\magnification=1200
%
%

%
%


\def\Serif{cmr}
\def\SerifBold{cmbx}
\def\SerifItalics{cmti}
\def\SerifSlanted{cmsl}
\def\SerifBoldItalics{cmbxti}
\def\SansSerif{cmss}
\def\SansSerifBold{cmssbx}
\def\SansSerifItalics{cmssi}
\def\SansSerifSlanted{cmssi}
\def\Math{cmmi}
\def\Symbols{cmsy}
\def\MathBold{cmmib}
\def\MoreSymbols{cmex}
\def\Typewriter{cmtt}
\def\Gothic{eufm}
\def\Double{msbm}

=            \Serif10            at 5pt
=        \SerifBold10        at 5pt
=     \SerifItalics10     at 5pt
=     \SerifSlanted10     at 5pt
= \SerifBoldItalics10 at 5pt
=        \SansSerif10        at 5pt
=    \SansSerifBold10    at 5pt
= \SansSerifItalics10 at 5pt
= \SansSerifSlanted10 at 5pt
=             \Math10             at 5pt
=         \MathBold10         at 5pt
=          \Symbols10          at 5pt
=      \MoreSymbols10      at 5pt
=       \Typewriter10       at 5pt
=           \Gothic10           at 5pt
=           \Double10           at 5pt

=           \Serif10            at 7pt
=       \SerifBold10        at 7pt
=    \SerifItalics10     at 7pt
=    \SerifSlanted10     at 7pt
=\SerifBoldItalics10 at 7pt
=       \SansSerif10        at 7pt
=   \SansSerifBold10    at 7pt
=\SansSerifItalics10 at 7pt
=\SansSerifSlanted10 at 7pt
=            \Math10             at 7pt
=        \MathBold10         at 7pt
=         \Symbols10          at 7pt
=     \MoreSymbols10      at 7pt
=      \Typewriter10       at 7pt
=          \Gothic10           at 7pt
=          \Double10           at 7pt

=           \Serif10            at 8pt
=       \SerifBold10        at 8pt
=    \SerifItalics10     at 8pt
=    \SerifSlanted10     at 8pt
=\SerifBoldItalics10 at 8pt
=       \SansSerif10        at 8pt
=   \SansSerifBold10    at 8pt
=\SansSerifItalics10 at 8pt
=\SansSerifSlanted10 at 8pt
=            \Math10             at 8pt
=        \MathBold10         at 8pt
=         \Symbols10          at 8pt
=     \MoreSymbols10      at 8pt
=      \Typewriter10       at 8pt
=          \Gothic10           at 8pt
=          \Double10           at 8pt

=             \Serif10            at 10pt
=         \SerifBold10        at 10pt
=      \SerifItalics10     at 10pt
=      \SerifSlanted10     at 10pt
=  \SerifBoldItalics10 at 10pt
=         \SansSerif10        at 10pt
=     \SansSerifBold10    at 10pt
=  \SansSerifItalics10 at 10pt
=  \SansSerifSlanted10 at 10pt
=              \Math10             at 10pt
=          \MathBold10         at 10pt
=           \Symbols10          at 10pt
=       \MoreSymbols10      at 10pt
=        \Typewriter10       at 10pt
=            \Gothic10           at 10pt
=            \Double10           at 10pt

=              \Serif10            at 12pt
=          \SerifBold10        at 12pt
=       \SerifItalics10     at 12pt
=       \SerifSlanted10     at 12pt
=   \SerifBoldItalics10 at 12pt
=          \SansSerif10        at 12pt
=      \SansSerifBold10    at 12pt
=   \SansSerifItalics10 at 12pt
=   \SansSerifSlanted10 at 12pt
=               \Math10             at 12pt
=           \MathBold10         at 12pt
=            \Symbols10          at 12pt
=        \MoreSymbols10      at 12pt
=         \Typewriter10       at 12pt
=             \Gothic10           at 12pt
=             \Double10           at 12pt

=            \Serif10            at 14pt
=        \SerifBold10        at 14pt
=     \SerifItalics10     at 14pt
=     \SerifSlanted10     at 14pt
= \SerifBoldItalics10 at 14pt
=        \SansSerif10        at 14pt
=    \SansSerifBold10    at 14pt
= \SansSerifSlanted10 at 14pt
= \SansSerifItalics10 at 14pt
=             \Math10             at 14pt
=         \MathBold10         at 14pt
=          \Symbols10          at 14pt
=      \MoreSymbols10      at 14pt
=       \Typewriter10       at 14pt
=           \Gothic10           at 14pt
=           \Double10           at 14pt

\def\NormalStyle{\parindent=5pt\parskip=3pt\normalbaselineskip=14pt%
\def\nt{\tenSerif}%
\def\rm{\fam0\tenSerif}%
\textfont0=\tenSerif\scriptfont0=\sevenSerif\scriptscriptfont0=\fiveSerif
\textfont1=\tenMath\scriptfont1=\sevenMath\scriptscriptfont1=\fiveMath
\textfont2=\tenSymbols\scriptfont2=\sevenSymbols\scriptscriptfont2=\fiveSymbols
\textfont3=\tenMoreSymbols\scriptfont3=\sevenMoreSymbols\scriptscriptfont3=\fiveMoreSymbols
\textfont\itfam=\tenSerifItalics\def\it{\fam\itfam\tenSerifItalics}%
\textfont\slfam=\tenSerifSlanted\def\sl{\fam\slfam\tenSerifSlanted}%
\textfont\ttfam=\tenTypewriter\def\tt{\fam\ttfam\tenTypewriter}%
\textfont\bffam=\tenSerifBold%
\def\bf{\fam\bffam\tenSerifBold}\scriptfont\bffam=\sevenSerifBold\scriptscriptfont\bffam=\fiveSerifBold%
\def\cal{\tenSymbols}%
\def\greekbold{\tenMathBold}%
\def\gothic{\tenGothic}%
\def\Bbb{\tenDouble}%
\def\LieFont{\tenSerifItalics}%
\nt\normalbaselines\baselineskip=14pt%
}

\def\TitleStyle{\parindent=0pt\parskip=0pt\normalbaselineskip=15pt%
\def\nt{\fourteenSansSerifBold}%
\def\rm{\fam0\fourteenSansSerifBold}%
\textfont0=\fourteenSansSerifBold\scriptfont0=\tenSansSerifBold\scriptscriptfont0=\eightSansSerifBold
\textfont1=\fourteenMath\scriptfont1=\tenMath\scriptscriptfont1=\eightMath
\textfont2=\fourteenSymbols\scriptfont2=\tenSymbols\scriptscriptfont2=\eightSymbols
\textfont3=\fourteenMoreSymbols\scriptfont3=\tenMoreSymbols\scriptscriptfont3=\eightMoreSymbols
\textfont\itfam=\fourteenSansSerifItalics\def\it{\fam\itfam\fourteenSansSerifItalics}%
\textfont\slfam=\fourteenSansSerifSlanted\def\sl{\fam\slfam\fourteenSerifSansSlanted}%
\textfont\ttfam=\fourteenTypewriter\def\tt{\fam\ttfam\fourteenTypewriter}%
\textfont\bffam=\fourteenSansSerif%
\def\bf{\fam\bffam\fourteenSansSerif}\scriptfont\bffam=\tenSansSerif\scriptscriptfont\bffam=\eightSansSerif%
\def\cal{\fourteenSymbols}%
\def\greekbold{\fourteenMathBold}%
\def\gothic{\fourteenGothic}%
\def\Bbb{\fourteenDouble}%
\def\LieFont{\fourteenSerifItalics}%
\nt\normalbaselines\baselineskip=15pt%
}

\def\PartStyle{\parindent=0pt\parskip=0pt\normalbaselineskip=15pt%
\def\nt{\fourteenSansSerifBold}%
\def\rm{\fam0\fourteenSansSerifBold}%
\textfont0=\fourteenSansSerifBold\scriptfont0=\tenSansSerifBold\scriptscriptfont0=\eightSansSerifBold
\textfont1=\fourteenMath\scriptfont1=\tenMath\scriptscriptfont1=\eightMath
\textfont2=\fourteenSymbols\scriptfont2=\tenSymbols\scriptscriptfont2=\eightSymbols
\textfont3=\fourteenMoreSymbols\scriptfont3=\tenMoreSymbols\scriptscriptfont3=\eightMoreSymbols
\textfont\itfam=\fourteenSansSerifItalics\def\it{\fam\itfam\fourteenSansSerifItalics}%
\textfont\slfam=\fourteenSansSerifSlanted\def\sl{\fam\slfam\fourteenSerifSansSlanted}%
\textfont\ttfam=\fourteenTypewriter\def\tt{\fam\ttfam\fourteenTypewriter}%
\textfont\bffam=\fourteenSansSerif%
\def\bf{\fam\bffam\fourteenSansSerif}\scriptfont\bffam=\tenSansSerif\scriptscriptfont\bffam=\eightSansSerif%
\def\cal{\fourteenSymbols}%
\def\greekbold{\fourteenMathBold}%
\def\gothic{\fourteenGothic}%
\def\Bbb{\fourteenDouble}%
\def\LieFont{\fourteenSerifItalics}%
\nt\normalbaselines\baselineskip=15pt%
}

\def\ChapterStyle{\parindent=0pt\parskip=0pt\normalbaselineskip=15pt%
\def\nt{\fourteenSansSerifBold}%
\def\rm{\fam0\fourteenSansSerifBold}%
\textfont0=\fourteenSansSerifBold\scriptfont0=\tenSansSerifBold\scriptscriptfont0=\eightSansSerifBold
\textfont1=\fourteenMath\scriptfont1=\tenMath\scriptscriptfont1=\eightMath
\textfont2=\fourteenSymbols\scriptfont2=\tenSymbols\scriptscriptfont2=\eightSymbols
\textfont3=\fourteenMoreSymbols\scriptfont3=\tenMoreSymbols\scriptscriptfont3=\eightMoreSymbols
\textfont\itfam=\fourteenSansSerifItalics\def\it{\fam\itfam\fourteenSansSerifItalics}%
\textfont\slfam=\fourteenSansSerifSlanted\def\sl{\fam\slfam\fourteenSerifSansSlanted}%
\textfont\ttfam=\fourteenTypewriter\def\tt{\fam\ttfam\fourteenTypewriter}%
\textfont\bffam=\fourteenSansSerif%
\def\bf{\fam\bffam\fourteenSansSerif}\scriptfont\bffam=\tenSansSerif\scriptscriptfont\bffam=\eightSansSerif%
\def\cal{\fourteenSymbols}%
\def\greekbold{\fourteenMathBold}%
\def\gothic{\fourteenGothic}%
\def\Bbb{\fourteenDouble}%
\def\LieFont{\fourteenSerifItalics}%
\nt\normalbaselines\baselineskip=15pt%
}

\def\SectionStyle{\parindent=0pt\parskip=0pt\normalbaselineskip=13pt%
\def\nt{\twelveSansSerifBold}%
\def\rm{\fam0\twelveSansSerifBold}%
\textfont0=\twelveSansSerifBold\scriptfont0=\eightSansSerifBold\scriptscriptfont0=\eightSansSerifBold
\textfont1=\twelveMath\scriptfont1=\eightMath\scriptscriptfont1=\eightMath
\textfont2=\twelveSymbols\scriptfont2=\eightSymbols\scriptscriptfont2=\eightSymbols
\textfont3=\twelveMoreSymbols\scriptfont3=\eightMoreSymbols\scriptscriptfont3=\eightMoreSymbols
\textfont\itfam=\twelveSansSerifItalics\def\it{\fam\itfam\twelveSansSerifItalics}%
\textfont\slfam=\twelveSansSerifSlanted\def\sl{\fam\slfam\twelveSerifSansSlanted}%
\textfont\ttfam=\twelveTypewriter\def\tt{\fam\ttfam\twelveTypewriter}%
\textfont\bffam=\twelveSansSerif%
\def\bf{\fam\bffam\twelveSansSerif}\scriptfont\bffam=\eightSansSerif\scriptscriptfont\bffam=\eightSansSerif%
\def\cal{\twelveSymbols}%
\def\bg{\twelveMathBold}%
\def\gothic{\twelveGothic}%
\def\Bbb{\twelveDouble}%
\def\LieFont{\twelveSerifItalics}%
\nt\normalbaselines\baselineskip=13pt%
}

\def\SubSectionStyle{\parindent=0pt\parskip=0pt\normalbaselineskip=13pt%
\def\nt{\twelveSansSerifItalics}%
\def\rm{\fam0\twelveSansSerifItalics}%
\textfont0=\twelveSansSerifItalics\scriptfont0=\eightSansSerifItalics\scriptscriptfont0=\eightSansSerifItalics%
\textfont1=\twelveMath\scriptfont1=\eightMath\scriptscriptfont1=\eightMath%
\textfont2=\twelveSymbols\scriptfont2=\eightSymbols\scriptscriptfont2=\eightSymbols%
\textfont3=\twelveMoreSymbols\scriptfont3=\eightMoreSymbols\scriptscriptfont3=\eightMoreSymbols%
\textfont\itfam=\twelveSansSerif\def\it{\fam\itfam\twelveSansSerif}%
\textfont\slfam=\twelveSansSerifSlanted\def\sl{\fam\slfam\twelveSerifSansSlanted}%
\textfont\ttfam=\twelveTypewriter\def\tt{\fam\ttfam\twelveTypewriter}%
\textfont\bffam=\twelveSansSerifBold%
\def\bf{\fam\bffam\twelveSansSerifBold}\scriptfont\bffam=\eightSansSerifBold\scriptscriptfont\bffam=\eightSansSerifBold%
\def\cal{\twelveSymbols}%
\def\greekbold{\twelveMathBold}%
\def\gothic{\twelveGothic}%
\def\Bbb{\twelveDouble}%
\def\LieFont{\twelveSerifItalics}%
\nt\normalbaselines\baselineskip=13pt%
}

\def\AuthorStyle{\parindent=0pt\parskip=0pt\normalbaselineskip=14pt%
\def\nt{\tenSerif}%
\def\rm{\fam0\tenSerif}%
\textfont0=\tenSerif\scriptfont0=\sevenSerif\scriptscriptfont0=\fiveSerif
\textfont1=\tenMath\scriptfont1=\sevenMath\scriptscriptfont1=\fiveMath
\textfont2=\tenSymbols\scriptfont2=\sevenSymbols\scriptscriptfont2=\fiveSymbols
\textfont3=\tenMoreSymbols\scriptfont3=\sevenMoreSymbols\scriptscriptfont3=\fiveMoreSymbols
\textfont\itfam=\tenSerifItalics\def\it{\fam\itfam\tenSerifItalics}%
\textfont\slfam=\tenSerifSlanted\def\sl{\fam\slfam\tenSerifSlanted}%
\textfont\ttfam=\tenTypewriter\def\tt{\fam\ttfam\tenTypewriter}%
\textfont\bffam=\tenSerifBold%
\def\bf{\fam\bffam\tenSerifBold}\scriptfont\bffam=\sevenSerifBold\scriptscriptfont\bffam=\fiveSerifBold%
\def\cal{\tenSymbols}%
\def\greekbold{\tenMathBold}%
\def\gothic{\tenGothic}%
\def\Bbb{\tenDouble}%
\def\LieFont{\tenSerifItalics}%
\nt\normalbaselines\baselineskip=14pt%
}

\def\AddressStyle{\parindent=0pt\parskip=0pt\normalbaselineskip=14pt%
\def\nt{\eightSerif}%
\def\rm{\fam0\eightSerif}%
\textfont0=\eightSerif\scriptfont0=\sevenSerif\scriptscriptfont0=\fiveSerif
\textfont1=\eightMath\scriptfont1=\sevenMath\scriptscriptfont1=\fiveMath
\textfont2=\eightSymbols\scriptfont2=\sevenSymbols\scriptscriptfont2=\fiveSymbols
\textfont3=\eightMoreSymbols\scriptfont3=\sevenMoreSymbols\scriptscriptfont3=\fiveMoreSymbols
\textfont\itfam=\eightSerifItalics\def\it{\fam\itfam\eightSerifItalics}%
\textfont\slfam=\eightSerifSlanted\def\sl{\fam\slfam\eightSerifSlanted}%
\textfont\ttfam=\eightTypewriter\def\tt{\fam\ttfam\eightTypewriter}%
\textfont\bffam=\eightSerifBold%
\def\bf{\fam\bffam\eightSerifBold}\scriptfont\bffam=\sevenSerifBold\scriptscriptfont\bffam=\fiveSerifBold%
\def\cal{\eightSymbols}%
\def\greekbold{\eightMathBold}%
\def\gothic{\eightGothic}%
\def\Bbb{\eightDouble}%
\def\LieFont{\eightSerifItalics}%
\nt\normalbaselines\baselineskip=14pt%
}

\def\AbstractStyle{\parindent=0pt\parskip=0pt\normalbaselineskip=12pt%
\def\nt{\eightSerif}%
\def\rm{\fam0\eightSerif}%
\textfont0=\eightSerif\scriptfont0=\sevenSerif\scriptscriptfont0=\fiveSerif
\textfont1=\eightMath\scriptfont1=\sevenMath\scriptscriptfont1=\fiveMath
\textfont2=\eightSymbols\scriptfont2=\sevenSymbols\scriptscriptfont2=\fiveSymbols
\textfont3=\eightMoreSymbols\scriptfont3=\sevenMoreSymbols\scriptscriptfont3=\fiveMoreSymbols
\textfont\itfam=\eightSerifItalics\def\it{\fam\itfam\eightSerifItalics}%
\textfont\slfam=\eightSerifSlanted\def\sl{\fam\slfam\eightSerifSlanted}%
\textfont\ttfam=\eightTypewriter\def\tt{\fam\ttfam\eightTypewriter}%
\textfont\bffam=\eightSerifBold%
\def\bf{\fam\bffam\eightSerifBold}\scriptfont\bffam=\sevenSerifBold\scriptscriptfont\bffam=\fiveSerifBold%
\def\cal{\eightSymbols}%
\def\greekbold{\eightMathBold}%
\def\gothic{\eightGothic}%
\def\Bbb{\eightDouble}%
\def\LieFont{\eightSerifItalics}%
\nt\normalbaselines\baselineskip=12pt%
}

\def\RefsStyle{\parindent=0pt\parskip=0pt%
\def\nt{\eightSerif}%
\def\rm{\fam0\eightSerif}%
\textfont0=\eightSerif\scriptfont0=\sevenSerif\scriptscriptfont0=\fiveSerif
\textfont1=\eightMath\scriptfont1=\sevenMath\scriptscriptfont1=\fiveMath
\textfont2=\eightSymbols\scriptfont2=\sevenSymbols\scriptscriptfont2=\fiveSymbols
\textfont3=\eightMoreSymbols\scriptfont3=\sevenMoreSymbols\scriptscriptfont3=\fiveMoreSymbols
\textfont\itfam=\eightSerifItalics\def\it{\fam\itfam\eightSerifItalics}%
\textfont\slfam=\eightSerifSlanted\def\sl{\fam\slfam\eightSerifSlanted}%
\textfont\ttfam=\eightTypewriter\def\tt{\fam\ttfam\eightTypewriter}%
\textfont\bffam=\eightSerifBold%
\def\bf{\fam\bffam\eightSerifBold}\scriptfont\bffam=\sevenSerifBold\scriptscriptfont\bffam=\fiveSerifBold%
\def\cal{\eightSymbols}%
\def\greekbold{\eightMathBold}%
\def\gothic{\eightGothic}%
\def\Bbb{\eightDouble}%
\def\LieFont{\eightSerifItalics}%
\nt\normalbaselines\baselineskip=10pt%
}

\def\ClaimStyle{\parindent=5pt\parskip=3pt\normalbaselineskip=14pt%
\def\nt{\tenSerifSlanted}%
\def\rm{\fam0\tenSerifSlanted}%
\textfont0=\tenSerifSlanted\scriptfont0=\sevenSerifSlanted\scriptscriptfont0=\fiveSerifSlanted
\textfont1=\tenMath\scriptfont1=\sevenMath\scriptscriptfont1=\fiveMath
\textfont2=\tenSymbols\scriptfont2=\sevenSymbols\scriptscriptfont2=\fiveSymbols
\textfont3=\tenMoreSymbols\scriptfont3=\sevenMoreSymbols\scriptscriptfont3=\fiveMoreSymbols
\textfont\itfam=\tenSerifItalics\def\it{\fam\itfam\tenSerifItalics}%
\textfont\slfam=\tenSerif\def\sl{\fam\slfam\tenSerif}%
\textfont\ttfam=\tenTypewriter\def\tt{\fam\ttfam\tenTypewriter}%
\textfont\bffam=\tenSerifBold%
\def\bf{\fam\bffam\tenSerifBold}\scriptfont\bffam=\sevenSerifBold\scriptscriptfont\bffam=\fiveSerifBold%
\def\cal{\tenSymbols}%
\def\greekbold{\tenMathBold}%
\def\gothic{\tenGothic}%
\def\Bbb{\tenDouble}%
\def\LieFont{\tenSerifItalics}%
\nt\normalbaselines\baselineskip=14pt%
}

\def\ProofStyle{\parindent=5pt\parskip=3pt\normalbaselineskip=14pt%
\def\nt{\tenSerifSlanted}%
\def\rm{\fam0\tenSerifSlanted}%
\textfont0=\tenSerif\scriptfont0=\sevenSerif\scriptscriptfont0=\fiveSerif
\textfont1=\tenMath\scriptfont1=\sevenMath\scriptscriptfont1=\fiveMath
\textfont2=\tenSymbols\scriptfont2=\sevenSymbols\scriptscriptfont2=\fiveSymbols
\textfont3=\tenMoreSymbols\scriptfont3=\sevenMoreSymbols\scriptscriptfont3=\fiveMoreSymbols
\textfont\itfam=\tenSerifItalics\def\it{\fam\itfam\tenSerifItalics}%
\textfont\slfam=\tenSerif\def\sl{\fam\slfam\tenSerif}%
\textfont\ttfam=\tenTypewriter\def\tt{\fam\ttfam\tenTypewriter}%
\textfont\bffam=\tenSerifBold%
\def\bf{\fam\bffam\tenSerifBold}\scriptfont\bffam=\sevenSerifBold\scriptscriptfont\bffam=\fiveSerifBold%
\def\cal{\tenSymbols}%
\def\greekbold{\tenMathBold}%
\def\gothic{\tenGothic}%
\def\Bbb{\tenDouble}%
\def\LieFont{\tenSerifItalics}%
\nt\normalbaselines\baselineskip=14pt%
}

%
%


\def\ModeYes{yes}
\def\ModeNo{no}

\def\ModeUndef{undefined}


\def\nx{\noexpand}
\def\ni{\noindent}
\def\newpage{\vfill\eject}

\def\ss{\vskip 5pt}
\def\ms{\vskip 10pt}
\def\bs{\vskip 20pt}

 \def\,{\mskip\thinmuskip}
 \def\!{\mskip-\thinmuskip}
 \def\>{\mskip\medmuskip}
 \def\;{\mskip\thickmuskip}

%
% ...
%

\def\refsModePost{post}
\def\refsModeAuto{auto}

\def\dbRefsSatusModeOk{ok}
\def\dbRefsSatusModeError{error}
\def\dbRefsSatusModeWarning{warning}


\newcount\BNUM
\BNUM=0

\def\refs{}

\def\SetModePost{\xdef\refsMode{\refsModePost}}         
\SetModePost

\def\dbRefsStatusOk{%
    \xdef\dbRefsStatus{\dbRefsSatusModeOk}%
    \xdef\dbRefsError{\ModeNo}%
    \xdef\dbRefsWarning{\ModeNo}%
    \xdef\dbRefsInfo{\ModeNo}%
}

\def\dbRefs{%
}

\def\dbRefsGet#1{%
    \xdef\found{N}\xdef\ikey{#1}\dbRefsStatusOk%
    \xdef\key{\ModeUndef}\xdef\tag{\ModeUndef}\xdef\tail{\ModeUndef}%
    \dbRefs%
}

\def\NextRefsTag{%
    \global\advance\BNUM by 1%
}
\def\ShowTag#1{{\bf [#1]}}

\def\dbRefsInsert#1#2{%
\dbRefsGet{#1}%
\if\found Y %
   \xdef\dbRefsStatus{\dbRefsSatusModeWarning}%
   \xdef\dbRefsWarning{record is already there}%
   \xdef\dbRefsInfo{record not inserted}%
\else%
   \toks2=\expandafter{\dbRefs}%
   \ifx\refsMode\refsModeAuto \NextRefsTag
    \xdef\dbRefs{%
    \the\toks2 \nx\xdef\nx\x{#1}%
    \nx\ifx\nx\ikey %
        \nx\x\nx\xdef\nx\found{Y}%
        \nx\xdef\nx\key{#1}%
        \nx\xdef\nx\tag{\the\BNUM}%
        \nx\xdef\nx\tail{#2}%
    \nx\fi}%
    \global\xdef\refs{\refs \ss\ni[\the\BNUM]\ #2\par}
   \fi%
   \ifx\refsMode\refsModePost
    \xdef\dbRefs{%
    \the\toks2 \nx\xdef\nx\x{#1}%
    \nx\ifx\nx\ikey %
        \nx\x\nx\xdef\nx\found{Y}%
        \nx\xdef\nx\key{#1}%
        \nx\xdef\nx\tag{\ModeUndef}%
        \nx\xdef\nx\tail{#2}%
    \nx\fi}%
   \fi%
\fi%
}

\def\dbRefsEdit#1#2#3{\dbRefsGet{#1}%
\if\found N
   \xdef\dbRefsStatus{\dbRefsSatusModeError}%
   \xdef\dbRefsError{record is not there}%
   \xdef\dbRefsInfo{record not edited}%
\else%
   \toks2=\expandafter{\dbRefs}%
   \xdef\dbRefs{\the\toks2%
   \nx\xdef\nx\x{#1}%
   \nx\ifx\nx\ikey\nx\x %
    \nx\xdef\nx\found{Y}%
    \nx\xdef\nx\key{#1}%
    \nx\xdef\nx\tag{#2}%
    \nx\xdef\nx\tail{#3}%
   \nx\fi}%
\fi%
}

\def\bib#1#2{\RefsStyle\dbRefsInsert{#1}{#2}%
    \ifx\dbRefsStatus\dbRefsSatusModeWarning %
        \message{^^J}%
        \message{WARNING: Reference [#1] is doubled.^^J}%
    \fi%
}

\def\ref#1{\dbRefsGet{#1}%
\ifx\found N %
  \message{^^J}%
  \message{ERROR: Reference [#1] unknown.^^J}%
  \ShowTag{??}%
\else%
    \ifx\tag\ModeUndef \NextRefsTag%
        \dbRefsEdit{#1}{\the\BNUM}{\tail}%
        \dbRefsGet{#1}%
        \global\xdef\refs{\refs \ss\ni [\tag]\ \tail\par}
    \fi
    \ShowTag{\tag}%
\fi%
}

\def\ShowBiblio{\bs\Ensure{\SectionEnsure}%
{\SectionStyle\ni References}%
{\RefsStyle\refs}%
}

\newcount\CHANGES
\CHANGES=0
\def\AuxFile{7}
\def\PreventDoubleOn{\xdef\PreventDoubleLabel{\ModeYes}}

\PreventDoubleOn

\def\StoreLabel#1#2{\xdef\itag{#2}
 \ifx\PreModeStatus\ModeNo %
   \message{^^J}%
   \errmessage{You can't use Check without starting with OpenPreMode (and finishing with ClosePreMode)^^J}%
 \else%
   \immediate\write\AuxFile{\nx\dbLabelPreInsert{#1}{\itag}}%
   \dbLabelGet{#1}%
   \ifx\itag\tag %
   \else%
    \global\advance\CHANGES by 1%
    \xdef\itag{(?.??)}%
    \fi%
   \fi%
}

\def\PreModeStatus{\ModeNo}

\def\ClosePreMode{\immediate\closeout\AuxFile%
  \ifnum\CHANGES=0%
    \message{^^J}%
    \message{**********************************^^J}%
    \message{**  NO CHANGES TO THE AuxFile  **^^J}%
    \message{**********************************^^J}%
 \else%
    \message{^^J}%
    \message{**************************************************^^J}%
    \message{**  PLAEASE TYPESET IT AGAIN (\the\CHANGES)  **^^J}%
    \errmessage{**************************************************^^ J}%
  \fi%
  \edef\PreModeStatus{\ModeNo}%
}

\def\dbLabelSatusModeOk{ok}

\def\dbLabelSatusModeWarning{warning}

\def\dbLabelStatusOk{%
    \xdef\dbLabelStatus{\dbLabelSatusModeOk}%
    \xdef\dbLabelError{\ModeNo}%
    \xdef\dbLabelWarning{\ModeNo}%
    \xdef\dbLabelInfo{\ModeNo}%
}

\def\dbLabel{%
}

\def\dbLabelGet#1{%
    \xdef\found{N}\xdef\ikey{#1}\dbLabelStatusOk%
    \xdef\key{\ModeUndef}\xdef\tag{\ModeUndef}\xdef\pre{\ModeUndef}%
    \dbLabel%
}

\def\ShowLabel#1{%
 \dbLabelGet{#1}%
 \ifx\tag \ModeUndef %
    \global\advance\CHANGES by 1%
    (?.??)%
 \else%
    \tag%
 \fi%
}

\def\dbLabelPreInsert#1#2{\dbLabelGet{#1}%
\if\found Y %
  \xdef\dbLabelStatus{\dbLabelSatusModeWarning}%
   \xdef\dbLabelWarning{Label is already there}%
   \xdef\dbLabelInfo{Label not inserted}%
   \message{^^J}%
   \errmessage{Double pre definition of label [#1]^^J}%
\else%
   \toks2=\expandafter{\dbLabel}%
    \xdef\dbLabel{%
    \the\toks2 \nx\xdef\nx\x{#1}%
    \nx\ifx\nx\ikey %
        \nx\x\nx\xdef\nx\found{Y}%
        \nx\xdef\nx\key{#1}%
        \nx\xdef\nx\tag{#2}%
        \nx\xdef\nx\pre{\ModeYes}%
    \nx\fi}%
\fi%
}

\def\dbLabelInsert#1#2{\dbLabelGet{#1}%
\def\itag{#2}%
\dbLabelGet{#1}%
\if\found Y %
    \ifx\tag\itag %
    \else%
       \ifx\PreventDoubleLabel\ModeYes %
        \message{^^J}%
        \errmessage{Double definition of label [#1]^^J}%
       \else%
        \message{^^J}%
        \message{Double definition of label [#1]^^J}%
       \fi%
    \fi%
   \xdef\dbLabelStatus{\dbLabelSatusModeWarning}%
   \xdef\dbLabelWarning{Label is already there}%
   \xdef\dbLabelInfo{Label not inserted}%
\else%
   \toks2=\expandafter{\dbLabel}%
    \xdef\dbLabel{%
    \the\toks2 \nx\xdef\nx\x{#1}%
    \nx\ifx\nx\ikey %
        \nx\x\nx\xdef\nx\found{Y}%
        \nx\xdef\nx\key{#1}%
        \nx\xdef\nx\tag{#2}%
        \nx\xdef\nx\pre{\ModeNo}%
    \nx\fi}%
\fi%
}


\newcount\PART
\newcount\CHAPTER
\newcount\SECTION
\newcount\SUBSECTION
\newcount\FNUMBER

\PART=0
\CHAPTER=0
\SECTION=0
\SUBSECTION=0
\FNUMBER=0

\def\LastPart{\ModeUndef}
\def\LastChapter{\ModeUndef}
\def\LastSection{\ModeUndef}
\def\LastSubSection{\ModeUndef}
\def\LastClaim{\ModeUndef}
\def\Last{\ModeUndef}

\newdimen\TOBOTTOM
\newdimen\LIMIT

\def\Ensure#1{\ \par\ \immediate\LIMIT=#1\immediate\TOBOTTOM=\the\pagegoal\advance\TOBOTTOM by -\pagetotal%
\ifdim\TOBOTTOM<\LIMIT\newpage \else%
\vskip-\parskip\vskip-\parskip\vskip-\baselineskip\fi}

\def\PartLabel{\the\PART}
\def\NewPart#1{\global\advance\PART by 1%
         \bs\ni{\PartStyle  Part \PartLabel:}
         \bs\ni{\PartStyle #1}\newpage%
         \CHAPTER=0\SECTION=0\SUBSECTION=0\FNUMBER=0%
         \gdef\Left{#1}%
         \global\edef\Last{\PartLabel}%
         \global\edef\LastPart{\PartLabel}%
         \global\edef\LastChapter{\ModeUndef}%
         \global\edef\LastSection{\ModeUndef}%
         \global\edef\LastSubSection{\ModeUndef}%
         \global\edef\LastClaim{\ModeUndef}}
\def\ChapterLabel{\the\CHAPTER}
\def\NewChapter#1{\global\advance\CHAPTER by 1%
         \bs\ni{\ChapterStyle  Chapter \ChapterLabel:}
         \bs\ni{\ChapterStyle #1}\ms%
         \SECTION=0\SUBSECTION=0\FNUMBER=0%
         \gdef\Left{#1}%
         \global\edef\Last{\ChapterLabel}%
         \global\edef\LastChapter{\ChapterLabel}%
         \global\edef\LastSection{\ModeUndef}%
         \global\edef\LastSubSection{\ModeUndef}%
         \global\edef\LastClaim{\ModeUndef}}
\def\SectionEnsure{3cm}
\def\SectionLabel{\the\SECTION}
\def\NewSection#1{\Ensure{\SectionEnsure}\global\advance\SECTION by 1%
         \bs\ni{\SectionStyle  \SectionLabel.\ #1}\ss%
         \SUBSECTION=0\FNUMBER=0%
         \gdef\Left{#1}%
         \global\edef\Last{\SectionLabel}%
         \global\edef\LastSection{\SectionLabel}%
         \global\edef\LastSubSection{\ModeUndef}%
         \global\edef\LastClaim{\ModeUndef}}
\def\NewAppendix#1#2{\Ensure{\SectionEnsure}\gdef\SectionLabel{#1}\global\advance\SECTION by 1%
         \bs\ni{\SectionStyle  Appendix \SectionLabel.\ #2}\ss%
         \SUBSECTION=0\FNUMBER=0%
         \gdef\Left{#2}%
         \global\edef\Last{\SectionLabel}%
         \global\edef\LastSection{\SectionLabel}%
         \global\edef\LastSubSection{\ModeUndef}%
         \global\edef\LastClaim{\ModeUndef}}
\def\Acknowledgements{\Ensure{\SectionEnsure}\gdef\SectionLabel{}%
         \bs\ni{\SectionStyle  Acknowledgments}\ss%
         \SECTION=0\SUBSECTION=0\FNUMBER=0%
         \gdef\Left{}%
         \global\edef\Last{\ModeUndef}%
         \global\edef\LastSection{\ModeUndef}%
         \global\edef\LastSubSection{\ModeUndef}%
         \global\edef\LastClaim{\ModeUndef}}
\def\SubSectionEnsure{2cm}
\def\SubSectionLabel{\ifnum\SECTION>0 \the\SECTION.\fi\the\SUBSECTION}
\def\NewSubSection#1{\Ensure{\SubSectionEnsure}\global\advance\SUBSECTION by 1%
         \ms\ni{\SubSectionStyle #1}\ss%
         \global\edef\Last{\SubSectionLabel}%
         \global\edef\LastSubSection{\SubSectionLabel}}
\def\SetNumberingModeN{\def\ClaimLabel{(\the\FNUMBER)}}
\def\SetNumberingModeSN{\def\ClaimLabel{(\ifnum\SECTION>0 \SectionLabel.\fi%
      \the\FNUMBER)}}
\def\SetNumberingModeCSN{\def\ClaimLabel{(\ifnum\CHAPTER>0 \the\CHAPTER.\fi%
      \ifnum\SECTION>0 \SectionLabel.\fi%
      \the\FNUMBER)}}

\def\NewClaim{\global\advance\FNUMBER by 1%
    \ClaimLabel%
    \global\edef\LastClaim{\ClaimLabel}%
    \global\edef\Last{\ClaimLabel}}
\def\fn{\eqno{\NewClaim}}
\def\fl#1{\fn\dbLabelInsert{#1}{\ClaimLabel}}
\def\fprel#1{\global\advance\FNUMBER by 1\StoreLabel{#1}{\ClaimLabel}\eqno{\itag}}

\def\cn{\NewClaim}
\def\cl#1{\global\advance\FNUMBER by 1\dbLabelInsert{#1}{\ClaimLabel}\ClaimLabel}
\def\cprel#1{\global\advance\FNUMBER by 1\StoreLabel{#1}{\ClaimLabel}\itag}

%
%


\def\al{\alpha}
\def\be{\beta}
\def\de{\delta}

\def\te{\theta}
\def\la{\lambda}
\def\ze{\zeta}
\def\om{\omega}
\def\si{\sigma}
\def\vp{\varphi}

\def\De{\Delta}
\def\Ga{\Gamma}

\def\La{\Lambda}
\def\Om{\Omega}

 
 \def\calU{{\hbox{\cal U}}}
 
 \def\calC{{\hbox{\cal C}}}
 \def\calP{{\hbox{\cal P}}}
 
 \def\calE{{\hbox{\cal E}}}


 \def\gotg{{\hbox{\gothic g}}}
 

 \def\one{{\hbox{\Bbb I}}}
 \def\A{{\hbox{\Bbb A}}}
 \def\R{{\hbox{\Bbb R}}}

 \def\E{{\hbox{\Bbb E}}}

 \def\F{{\hbox{\Bbb F}}}
 
 \def\R{{\hbox{\Bbb R}}}

\def\Tr{{\hbox{Tr}}}

\def\Aut{{\hbox{Aut}}}

\def\ad{{\hbox{ad}}}
\def\Ad{{\hbox{Ad}}}
\def\AD{{\hbox{\bf Ad}}}

\def\SO{{\hbox{SO}}}

\def\GL{{\hbox{GL}}}

\def\d{{\hbox{d}}}

\def\f{{\hbox{\bf f}}}
\def\Tf{{\hbox{\bf Tf}}}
\def\tf{T\f}

\def\ip{\hbox to4pt{\leaders\hrule height0.3pt\hfill}\vbox to8pt{\leaders\vrule width0.3pt\vfill}\kern 2pt} 
\def\QDE{\hfill\hbox{\ }\vrule height4pt width4pt depth0pt}
\def\del{\partial}
\def\na{\nabla}

\def\Lie{\hbox{\LieFont \$}}

\def\arr{\rightarrow}

\def\then{\Rightarrow}
\def\semidirect{\hbox{\Bbb \char111}}
\def\binomial#1#2{\left(\matrix{#1\cr #2\cr}\right)}

%
%

\def\PROPERTYn{\ClaimStyle\ni{\bf Property \cn: }}

\def\ENDPROPERTY{\NormalStyle}

\def\COROLLARYn{\ClaimStyle\ni{\bf Corollary \cn: }}

\def\ENDCOROLLARY{\NormalStyle}

\def\THEOREMl#1{\ClaimStyle\ni{\bf Theorem \cl{#1}: }}
\def\ENDTHEOREM{\NormalStyle}

\def\PROOF{\ProofStyle\ni{\bf Proof: }}
\def\ENDPROOF{\hfill\QDE\NormalStyle}

\NormalStyle
\SetNumberingModeSN
\PreventDoubleOn

\long\def\title#1{\centerline{\TitleStyle\ni#1}}
\long\def\author#1{\bs\centerline{\AuthorStyle by {\it #1}}}

\long\def\address#1{\ss\centerline{\AddressStyle #1}\par}
\long\def\moreaddress#1{\centerline{\AddressStyle #1}\par}
\def\abstract{\bs\leftskip 1cm\rightskip .5cm\AbstractStyle{\bf \ni Abstract:}\ }
\def\endabstract{\par\leftskip 0cm\rightskip 0cm\NormalStyle\ss}

\bib{Libro}{L. Fatibene, M. Francaviglia,
{\it Natural and Gauge Natural Formalism for Classical Field Theories},
Kluwer Academic Publishers, (Dordrecht, 2003), xxii
}

\bib{Kolar}{I.\ Kol{\'a}{\v r}, P.W.\ Michor, J.\ Slov{\'a}k,
{\it Natural Operations in Differential Geometry},
Springer--Verlag, (New York, 1993)
}

\bib{KobaNu}{S. Kobayashi, K. Numizu,
{\it Foundations of Differential Geometry. Volume I},
John Wiley \& Sons, Inc. Interscience Division, (New York,  1963)
}

\bib{Augmented}{L. Fatibene, M. Ferraris, M. Francaviglia,
{\it  Int. J. Geom. Methods Mod. Phys.}, v.2, N3, (2005)
 }

\bib{Root}{M. Ferraris, M. Francaviglia,
in: {\it Mechanics, Analysis and Geometry: 200 Years after Lagrange},
Editor: M. Francaviglia; Elsevier Science Publishers B.V., (Amsterdam, 1991)
}

\bib{Cavalese}{M. Ferraris, M. Francaviglia,
in: {\it 8th Italian Conference on General Relativity and Gravitational Physics}, Cavalese (Trento),
August 30--September 3, World Scientific, (Singapore, 1988)
}

\bib{AFR}{G. Allemandi, M. Francaviglia, M. Raiteri,
{\it Classical Quantum Gravity} {\bf 20}(3), 483, (2003); 
gr-qc/0211098
}

\bib{BF}{A. Borowiec, M. Francaviglia,
in: {\it Vietri sul Mare 2004, General Relativity and Gravitational Physics},
AIP Conf.~Proc. {\bf 751},
165, (2005);
math-ph/0506068
}

\bib{BFa}{D. Birmingham, M. Blau, M. Rakowski, G. Thompson,
{\it Phys. Rep.}, {\bf 209}, 129,  (1991)
}

\bib{BFb}{A. S. Cattaneo, P. Cotta-Ramusino, J. Fr\"ohlich, M. Martellini,
{\it J. Math. Phys.}, {\bf 36} 6137,  (1995)
}

\bib{BFc}{M. Montesinos,
{\it Class. Quantum Grav.}, {\bf 20}, 3569,  (2003)
}

\bib{BFd}{G. Sardanashvily,
{\it Int. J. Geom. Meth. Mod. Phys.}, {\bf 1}, 4 (2004)
}

\bib{BFF}{A. Borowiec, M. Ferraris, M. Francaviglia,
{\it J.~Phys.~A}, {\bf 31}(44), 8823, (1998)
}

\bib{BFFa}{A. Borowiec, M. Ferraris, M. Francaviglia,
{\it J.~Phys.~A},  {\bf 36}(10), 2589, (2003)
}

\bib{BFFP}{A. Borowiec, M. Ferraris, M. Francaviglia, M. Palese,
{\it Universitatis Iagellonicae Acta Mathematica}, {\bf XLI} 319, (2003)
}

\bib{Tapia}{M. Ferraris, M.Francaviglia, V. Tapia,
{\it J. Phys. A} {\bf26}(2), 433, (1993)
}

\bib{Silvio}{L. Fatibene, M.\ Ferraris, M.\ Francaviglia, S. Mercadante,
{\it Int. J. Geom. Methods Mod. Phys.}, {\bf 2}(5), (2005)
}

\bib{Ray}{L.\ Fatibene, M.\ Ferraris, M.\ Francaviglia, R.G.\ McLenaghan
{\it J. Math. Phys.} {\bf 43}(6), 3147, (2002)
}

\bib{Sard1}{G. Sardanashvily,
{\it Energy-momentum conservation laws in higher-dimensional Chern-Simons models};
hep-th/0303148
}

\bib{Sard2}{G. Sardanashvily,
{\it Gauge conservation laws in higher-dimensional Chern-Simons models};
hep-th/0303059
}

\bib{GMS}{G.Giachetta, L.Mangiarotti, G. Sardanashvily,
Mod.~Phys.~Lett.~A{\bf 18} 2645
(2003); math-ph/0310067
}

\bib{Lovelock}{G.Allemandi, M.Francaviglia, M.Raiteri,
Class.~Quant.~Grav.~{\bf 20},
5103, (2003);
gr-qc/0308019
}

\bib{Stasheff}{R. Fulp, T. Lada, J. Stasheff,
in: {\it Proceedings of 22nd Winter School on Geometry and Physics}, Srni, Czech Republic, 12-19 Jan 2002
(to appear);
math.qa/0204079.
}

\bib{CS1}{S. Chern, J. Simons,
Proc. Nat. Acad. Sci. USA {\bf 68}(4), 791, (1971)
}

\bib{CS2}{S. Chern, J. Simons,
Ann. Math. {\bf 99}, 48, (1974)
}

\bib{Montesinos}{M. Mondrag\'on, M. Montesinos,
{\it Covariant canonical formalism for $4$-dimensional BF theory},
J. Math. Phys. (in print); gr-qc/0402041
}

\bib{Izaurieta}{F. Izaurieta, E. Rodriguez, P. Salgado,
{\it On Transgression Forms and Chern--Simons (Super)gravity}; 
hep-th/0512014
}

\def\AppendixTapia{A}
\def\AppendixFE{B}
\topskip=50pt
\ \vskip 20pt

\title{Covariant Lagrangian Formulation of Chern-Simons and BF Theories}

\author{A.\ Borowiec$^1$, L.\ Fatibene$^{2, 3}$, M.\ Ferraris$^2$, M.\ Francaviglia$^{2, 3, 4}$}

\address{$^1$Institute of Theoretical Physics, University of Wroc\l aw (Poland)}

\moreaddress{$^2$Dipartimento di Matematica, Universit\`a degli Studi di Torino (Italy)}

\moreaddress{$^3$INFN, Sezione di Torino, Iniziativa Specifica NA12 (Italy)}

\moreaddress{$^4$ESG, Universit\`a della Calabria (Italy)}

\abstract
We investigate the covariant formulation of Chern-Simons theories in a general odd dimension   which can be obtained by introducing a vacuum connection field as a reference.
Field equations, N\"other currents and superpotentials are computed so that results are easily compared with the well-known results in dimension $3$.
Finally we use this covariant formulation of Chern-Simons theories to investigate their relation with
topological BF theories.
\endabstract

\NewSection{Introduction}

Chern-Simons theories are, in their original formulation, global field theories defined by
sheafs of local Lagrangians depending on some (principal) connection. The local Lagrangians on two intersecting patches differ in the patch intersection by a local divergence which depends on the gauge choice.
Despite these local divergences prevent the local Lagrangians to
glue together and form a global Lagrangian, the local Lagrangians altogether define in each patch the same gauge-covariant field equations which are thence the appropriate restrictions of a single globally well-defined field equation.

However, a globalization procedure is usually defined only after another connection has been fixed.
This allows one to define a global Lagrangian depending on two connections (together with their curvatures)
which is gauge covariant and induces the same (global) field equations of the original local Lagrangian sheaf.

Conservation laws are particularly tricky when local Lagrangians are considered. Due to the local divergence indetermination, the local Lagrangians are not gauge invariant. They are just locally invariant modulo local divergences.
In this situation the application of N\"other theorem is tricky and the definition of N\"other currents needs some extra care.
On the contrary, when a global covariant Lagrangian is used these problems disappear altogether and superpotentials can be computed as the theory is now set as a global gauge natural theory (see \ref{Libro}, \ref{Sard1}, \ref{Sard2}, \ref{GMS}).

The situation was preliminary discussed in detail in dimension $3$ (see \ref{BFF}, \ref{BFFa}, \ref{BFFP} and references quoted therein). It will be discussed here in any arbitrary odd dimension.
As it should be expected, the explicit expression for a covariant global Lagrangian grows in complication with dimension;
these expressions are however interesting since they present a family of global topological theories.
In this paper we shall explicitly consider the case of Chern-Simons Lagrangians for gauge theories;
we shall not deal with Chern-Simons gravitational models, which will be treated elsewhere (see \ref{Lovelock}).

\NewSection{Notation}

Let $\calP=(P, M, \pi, G)$ be a principal bundle.
Even if the notation $\calP$ and $P$ should be in principle reserved to denote respectively the bundle and its total space, we shall, by an abuse of language (and for simplicity), consider them as being equivalent.
Let $\gotg$ denote the Lie algebra of the structure group $G$ and $T_A$ be a basis of $\gotg$.
If the group is semisimple then $T_A$ is usually chosen to be orthonormal with respect to
the Cartan-Killing metric of $G$.

We shall hereafter denote by $\ad:G\times G\arr G: (g, h)\mapsto g\cdot h\cdot g^{-1}$ the adjoint action of the group onto itself,
by $T_e\ad\equiv\Ad:G\times \gotg\arr \gotg: (g, T_A)\arr \Ad_A^B(g) T_B$ the induced adjoint action  of the group onto the Lie algebra, and
by $\AD:\gotg\times \gotg\arr \gotg: [T_A, T_B]\mapsto c_{AB}{}^C T_C$ the induced adjoint action of the Lie algebra onto itself.
The constants $c_{AB}{}^C$ are the {\it structure constants} of the group $G$; they depend of course on the basis $T_A$.

Fibered coordinates on $\calP$ are denoted by $(x^\mu, g^a)$;
the Lie algebra is identified with the tangent space $T_eG$ to the group at the identity $e\in G$.
Accordingly, we set $T_A= T_A^a\del_a\vert_{e}$
(where $\del_a\vert_{e}$ denotes the basis of tangent vectors to $G$ induced by the coordinates chosen around the identity $e$).
We denote by $R_g: P\arr P: p\mapsto p\cdot g$ the canonical right action
of the group $G$ on the principal bundle $P$ and by
$T_eR_g: \gotg\arr \gotg$ the induced right action on the algebra
given by $TR_g(T_A)=: R_A^B(g) T_B=: R_A^a(g)\del_a\vert_{e} $.
Let $\rho_A$ denote a right invariant pointwise basis of vertical vectors corresponding to
the basis $T_A$. One has $ \rho_A=R_A^a(g) \del_a$.

As it is standard in the gauge natural framework (see \ref{Libro}) $W^{(1,1)}G$ will denote the semidirect product $W^{(1,1)}G= \GL(m) \semidirect J^1G$ and the fibered product
$W^{(1,1)}\calP= L(M) \times_M J^1\calP$ denotes the gauge natural prolongation of order $(1,1)$ of the principal bundle $\calP$ (see \ref{Libro}, \ref{Kolar}).
The bundle $W^{(1,1)}\calP$ is a principal bundle with $W^{(1,1)}G$ as structure group.
If $(x^\mu, V^\mu_a)$ are fibered coordinates of $L(M)$,
then $(x^\mu, V^\mu_a, g^a, g^a_\mu)$ are fibered coordinates on $W^{(1,1)}\calP$.

Let us now set $\A= (\R^m)^\ast\otimes \gotg$ (considered as an affine space with local coordinates $A^A_\mu$) and let us consider the following affine action on the left:
$$
\la: W^{(1,1)}G\times \A\arr \A: (J, g, dg, A)\mapsto (\Ad_B^A(g)
A^B_\nu + \bar R^A_b(g) g^b_\nu)\bar J^\nu_\mu
\fl{ConnectionTransformationEQ}$$ where the bars denote
matrix inversion. The associated bundle $\calC(P) := W^{(1,1)}\calP \times_\la
\A$ is known to be canonically isomorphic to $J^1\calP/G$. Its
global sections are in one-to-one canonical correspondence with
principal connections on $\calP$, which we generically denote by $A$. We stress that, by definition,
$\calC(P)$ is a bundle over $M$, despite principal connections in
mathematical literature usually {\it live} on $P$ and their corresponding objects on $M$
are obtained by gauge fixing. Our representation for principal
connections is instead done in terms of sections of a bundle over $M$. Nevertheless,
it is manifestly global and gauge covariant and it does not rely on any gauge
fixing.

Given a principal connection on $\calP$ under its (local) form $\om=\d x^\mu\otimes (\del_\mu- A^A_\mu(x)\> \rho_A)$,  its {\it curvature} is given by
$$
F=\hbox{$1\over 2$} F^A_{\mu\nu} T_A\otimes \d x^\mu\land \d x^\nu,
\qquad
F^A_{\mu\nu}= \d_\mu A^A_\nu - \d_\nu A^A_\mu + c_{BC}{}^A A^B_\mu A^C_\nu
\fl{CurvatureEQ}$$

Let us now denote  by $\La^k(P; \gotg)$ the bundle 
of $k$-forms on $M$ with values in the bundle $\La^0(P; \gotg)$, for
any positive integer $k$. For $k=0$ we obtain $\La^0(P; \gotg)$, i.e.\ the
bundle of (infinitesimal) gauge transformations. 
The bundle $\La^k(P; \gotg)$ is associated to $L(M)\times_M P$
through the following action on $V=\La^k(\R^m)\otimes \gotg$:
$$
l: \GL(m) \times G \times V \mapsto V: (J, g, \om)\mapsto
\Ad^A_B(g)\>\om^B_{\rho\si\dots \te}\bar J_\mu^\rho  \bar J_\nu^\si
\dots \bar J_\la^\te 
\fl{CurvatureTransformationLawsEQ}$$ 
The bundles $\La^k(P; \gotg)$ are by construction gauge natural bundles
of order $(1,0)$. $\La^1(P; \gotg)$
is the vector bundle which the affine connection bundle $\calC(P)$
is modeled on; the difference of two connections $A$ and $\bar A$ is a section of
$\La^1(P; \gotg)$, which we denote by $\al=\al(A, \bar A)= A-\bar A$. 
The curvature of a connection is a section of $\La^2(P; \gotg)$.
Equation $\ShowLabel{CurvatureEQ}$ is in fact the local expression of a
global bundle morphism $\hbox{\bf F}:J^1\calC(P)\arr \La^2(P;
\gotg)$.

Let us remark that, by construction, our bundles $\La^k(P;
\gotg)\cong\La^k(M)\otimes\La^0(P; \gotg)$ are bundles over $M$, not
to be confused with the bundles $\La^k(P)\otimes\gotg$ of
$\gotg$-valued $k$-forms on $P$ (here $\gotg$ can be identified with
the trivial bundle $P\times\gotg$) used frequently in literature
(e.g. \ref{KobaNu}) and often denoted in a similar way (e.g.
\ref{Kolar}) which, on the contrary, are defined as bundles over $P$.
In particular, $\La^0(P)\otimes\gotg\cong P\times\gotg$, which coincides with
the vertical bundle of the principal bundle $P$, is always a trivial
vector bundle over $P$, while our $\La^0(P; \gotg)$ is, in general,
a nontrivial bundle over $M$. More generally, there is a one-to-one
correspondence between usual sections of $\La^k(P; \gotg)$
and the so-called {\it tensorial} $\gotg-$valued
$k-$forms on $P$, i.e. {\it tensorial} sections of $\La^k(P)\otimes\gotg$ \ref{KobaNu}. 
We stress that any standard connection $1$-form is a
vertical and $\Ad-$invariant section of the bundle
$\La^1(P)\otimes\gotg$ over $P$ but it is not a global section of
the bundle $\La^1(P; \gotg)$ over $M$. To obtain the gauge fields
often used in Physics (see \ref{KobaNu}) which are {\it living} on $M$ one
needs to pull-back the connection $1$-form along a section of $P$ which exists globally 
if and only if $P$ is trivial. 
This procedure is usually called a {\it gauge
fixing} and we shall try to avoid it in order to work with objects
which are already global in the most general situation.

The wedge product needs to be extended to forms with values in the Lie algebra (owing to the fact that
the product of components needs to be replaced by Lie product, which is the only product in a
generic Lie algebra).
We define, e.g.\ on $1$-forms
$$
[\te, \la] := c^A{}_{BC} \>\te^B_\mu \la^C_\nu\> T_A\otimes \d x^\mu\land \d x^\nu
\fl{ExteriorCovariantDifferential}$$
and similar more complicated expression hold for $k$-forms with $k>1$.
Notice in $\ShowLabel{ExteriorCovariantDifferential}$ the absence of combinatorial coefficients in the resulting $2$-form.
Let us also set for a $k$-form $\te$:
$$
(\te^2):= \hbox{$1\over (2k)!$} [\te, \te]
\fn$$
which we stress to be non-zero for odd $k$.
Hereafter we shall use this notation only in the case $k=1$; e.g.~when using $(\al^2)$.

The {\it exterior differential} of a $k$-form  $\te=\hbox{$1\over k!$} \te^A_{\mu_1\mu_2\dots \mu_k}\>
 T_A\otimes\d x^{\mu_1}\land\d x^{\mu_2}\land\dots\land \d x^{\mu_k}$ in $\La^k(P; \gotg)$ is defined as
$$
\d \te=\hbox{$1\over k!$} \d_\mu \te^A_{\mu_1\mu_2\dots \mu_k}\>
T_A\otimes\d x^\mu\land \d x^{\mu_1}\land\d x^{\mu_2}\land\dots\land \d x^{\mu_k}
\fn$$
It is not gauge covariant; hence, once a connection $A$ is fixed on $P$, the {\it exterior covariant differential} can be defined as
$$
\na \te= \d \te + [A, \te]
\fl{CovariantDerivativeofAFormAlgebraEQ}$$
Of course a connection $A$ is not a section of the bundle $\La^1(P;
\gotg)$ over $M$ as eq. \ShowLabel{CovariantDerivativeofAFormAlgebraEQ} should suggest (or, equivalently, connections are not {\it tensorial}; see \ref{KobaNu}).
Here a standard abuse of language is used to suitably identify a connection $A$ with the (local) $1$-form
$\al=A-\bar A$ once, in a trivialization, a {\it trivial} background connection $\bar A=0$ is set.
Hereafter and systematically in all local expressions we shall locally represent a connection by this
section $\al_0=\al(A, \bar 0)=A-\bar 0=A$ of $\La^1(P; \gotg)$; accordingly we have the well-known expression
$$
F= \d A + \hbox{$1\over 2$} [A, A]
\fl{CurvatureExpressionsWithForms}$$
for the curvature.
By this abuse of language we shall earn a compact way for writing local expressions, even though the formulae so obtained
are not completely meaningful from a global point of view. For example, despite the curvature
\ShowLabel{CurvatureExpressionsWithForms} is globally well defined the terms $\d A $ and $\hbox{$1\over 2$} [A, A]$ are not separately global.

Let now $\Phi:P\arr P$ be a principal automorphism projecting over a diffeomorphism
$\phi:M\arr M$. Its local expression is of the form
$$
\cases{
x'= \phi(x)\cr
g' = \vp(x)\cdot g\cr
}
\fn$$
where $\vp(x)$ is a group isomorphism in $G$.
The infinitesimal generator of a $1$-parameter subgroup of principal automorphisms
is a projectable right invariant vector field, i.e.:
$$
\Xi= \xi^\mu(x)\del_\mu + \xi^A(x)\rho_A
\fn$$

Principal automorphisms are also known as {\it gauge transformations}.
They canonically induce automorphisms on any gauge natural bundle associated to $\calP$,
in particular on $\calC(P)$ and all bundles $\La^k(P; \gotg)$; the canonically induced vector fields
on $\calC(P)$ and $\La^2(P; \gotg)$ are:
$$
\calC(\Xi)= \xi^\mu(x)\del_\mu + \xi^A_\mu \del_A^\mu,
\qquad
\xi^A_\mu=  \d_\mu \xi^A +c_{BE}{}^A A^B_\mu \xi^E - \d_\mu\xi^\rho  A^A_\rho
\fn$$
and:
$$
\La^2(\Xi; \gotg)=\xi^\mu(x)\del_\mu + \xi^A_{\mu\nu} \del_A^{\mu\nu},
\qquad
\xi^A_{\mu\nu}= c_{BC}{}^A\xi^B F^C_{\mu\nu}
-\d_\mu\xi^\rho F^A_{\rho\nu} - \d_\nu\xi^\rho F^A_{\mu\rho}
\fn$$
respectively.

Hence we can define the Lie derivative of gauge natural objects with respect to
an infinitesimal gauge generator $\Xi$.
We easily obtain the following specific rules:
$$
\Lie_\Xi A^A_\mu= \xi^\la F^A_{\la\mu} + \na_\mu (\xi^A_{(V)})
\fn$$
and
$$
\Lie_\Xi F^A_{\mu\nu}= \xi^\rho \na_\rho (F^A_{\mu\nu})
+ \na_\mu\xi^\rho F^A_{\rho\nu}+ \na_\nu\xi^\rho F^A_{\mu\rho}
- c_{BC}{}^A \xi^B_V F^C_{\mu\nu}
\fn$$
where $\na_\mu$ denotes covariant derivative with respect to the connection $A^A_\mu$
(as well as to a symmetric base connection $\Ga^\al_{\be\mu}$)
and $\xi^A_{(V)}= \xi^A+ A^A_\mu \xi^\mu$ denotes the vertical part of $\Xi$.
Let us remark that the contribution of the symmetric base connection will be systematically cancelled out
hereafter by symmetry reasons (e.g. in the exterior covariant differential below).
More precisely we have:
$$
\na_\mu(\xi^A_{(V)})= \d_\mu \xi^A_{(V)} + c_{BC}{}^A A^B_\mu \xi^C_{(V)}
\fn$$
and
$$
\na_\rho(F^A_{\mu\nu})= \d_\rho F^A_{\mu\nu}
- \Ga^\la_{\mu\rho} F^A_{\la\nu}
- \Ga^\la_{\nu\rho} F^A_{\mu\la}
+ c_{BC}{}^A A^B_\rho F^C_{\mu\nu}
\fl{nablaFEQ}$$

Using $\ShowLabel{CurvatureEQ}$ and $\ShowLabel{nablaFEQ}$ one easily obtains the so-called {\it Bianchi identities}:
$$
\eqalign{
\na_{[\rho} F^A_{\mu\nu]}=&
\d_{[\rho} F^A_{\mu\nu]}
+ c_{BC}{}^A A^B_{[\rho} F^C_{\mu\nu]}=
( c_{[DE}{}^C c_{B]C}{}^A) A^B_{\rho} A^D_\mu A^E_{\nu}
\equiv 0 \cr
}
\fn$$
where Jacobi identities in the Lie algebra have been implicitly used.
An intrixic expression for Bianchi identies is obtained by means of the so-called
{\it exterior covariant differential} with respect to the connection $A^A_\mu$ (see \ref{Libro}, \ref{Kolar}, \ref{KobaNu}):
$$
\na F:= \hbox{$1\over 2$} \na_{[\rho} F^A_{\mu\nu]}\> T_A\otimes \d x^\rho\land\d x^\mu\land \d x^\nu
\equiv0
\fn$$

\NewSection{Invariant Polynomials}

A {\it symmetric polynomial} on a vector space $V$ is a symmetric $k$-linear form, $k$ being any positive integer.
We denote by $S^k(V)=\{ f:V^k\arr \R: f\>\> \hbox{symmetric and $k$-linear}\}$.

If $V$ happens to be the Lie algebra $V\equiv \gotg$ of a Lie group $G$, then an {\it invariant polynomial}
is a symmetric polynomial which is also $\Ad$-invariant,
i.e.\ $\forall g\in G$ and $\forall\xi_1, \xi_2, \dots, \xi_k\in \gotg$:
$$
f(\Ad_g(\xi_1),\Ad_g(\xi_2), \dots, \Ad_g(\xi_k))=f(\xi_1, \xi_2, \dots, \xi_k)
\fn$$
The set of all invariant polynomials of degree $k$ is denoted by $I^k(\gotg)$.
We set $\deg(f):=k$ for each $f\in I^k(\gotg)$.

The prototypes of invariant polynomials are the power-trace polynomials.
By using the adjoint action $\AD$ an element $\xi$ of $\gotg$ can be identified with an
endomorphism $\AD(\xi):\gotg\arr \gotg: \ze\mapsto [\xi,\ze]=\xi^A\ze^B c_{AB}{}^C T_C$.
The matrix representing such an endomorphism is:
$$
\AD^C_B(\xi)= \xi^A c_{AB}{}^C
\fn$$
We can in particular consider $f_k=\Tr(\AD^k)$; for example:
$$
\eqalign{
&f_2(\xi_1, \xi_2)= \xi_1^A\xi_2^B c_{AE}{}^D  c_{BD}{}^E \in I^2(\gotg)\cr
&f_3(\xi_1, \xi_2, \xi_3)= \xi_1^A\xi_2^B\xi_3^C c_{AE}{}^D  c_{BF}{}^E c_{CD}{}^F
\in I^3(\gotg)\cr
}
$$
and so on.

These polynomials are trivially symmetric and $\Ad$-invariant;
one can easily check that in specific examples  they are not trivially vanishing (e.g.~when $G=\SO(3)$).
Before entering in details let us first briefly summarize the main points of the construction leading to Chern-Simons theories  (see \ref{CS1}, \ref{CS2}).

Once a connection $A$ has been fixed on $P$ there exists a canonical prescription that
to any invariant polynomial $f\in I^k(\gotg)$ associates a closed horizontal $2k$-form on $\calP$ denoted by $f(F^k)\in \La^{2k}(P)$.
This forms is {\it transgressive}, i.e.\ it is the pull-back along the projection $\pi:P\arr M$ of a $2k$-form $\f(F^k)$ on $M$ (see \ref{Izaurieta}).
Despite the form $\f(F^k)$ depends on the connection $A$ one can check that the cohomology class
it identifies in $H^{2k}(M,\R)$ is independent of $A$. In fact,
if any two connections $A$ and $\bar A$ are fixed on $P$ one is then able to {\it constructively} single out a
representative $\tf(A, \bar A)$ of this cohomology class.
This is a global $(2k-1)$-form on $M$.
Hence a local potential $\tf(A)=\tf(A, 0)$ (which is global on trivial bundles) is singled out on $M$ for the form $\f(F^k)$.

If the dimension of $M$ is an odd integer $m=2k-1$, $k:=\deg(f)$, then $\tf(A, \bar A)$ is a global closed $(2k-1)$-form and the local forms $\tf(A)$ can be used as a sheaf of local Lagrangians; we shall prove that these {\it local} Lagrangians induce {\it global} field equations.

We remark that $\tf(A, \bar A)$ will be used as a convenient globalization of the local Lagrangians
$\tf(A)$.
As it often happens, globality is obtained by introducing a background (or, using a better terminology introduced in \ref{Augmented}, a {\it reference vacuum state}) and mimicking the covariant first order Lagrangian known to exist for General Relativity (see \ref{Augmented}, \ref{Root}, \ref{Cavalese}).

Let us also stress that the bundle level $P$ is essential to the construction.
Once again (as, e.g., for superpotentials) objects have no canonical representative at the base level, though canonical representatives can be defined at the bundle level and eventually be pulled back on the base manifold.
For example, in the case of interest $m=2k-1$, the form $f(F^k)$ on $M$ identically vanishes due to dimensional reasons. The potential $\tf(A, \bar A)$ we shall build is not zero (as one could expect by working in the base manifold). It is important to regard $\tf(A, \bar A)$ as a map defined in general for all $k$, all $m$ and all topologies of $P$, to be later specialized to particular cases (even when for other reasons other {\it canonical representatives} exist).

Let us hence begin to define the correspondence between invariant polynomials and bundle forms in the general case ($\dim(M)=m$ and $k$ any pair of positive integers).
Let $f\in I^k(\gotg)$ be an invariant polynomial of degree $k$;
let us denote by $f_{A_1A_2\dots A_k}:=f(T_{A_1}, T_{A_2},\dots, T_{A_k})$ its symmetric  coefficients.
Since the curvature $F\in \La^2(P;\gotg)$ is a $\gotg$-valued $2$-form on $P$
we set
$$
f(F^k)= \hbox{$1\over 2^k$}f_{A_1A_2\dots A_k} F^{A_1}_{\mu_1\nu_1} F^{A_2}_{\mu_2\nu_2}\dots F^{A_k}_{\mu_k\nu_k}
\d x^{\mu_1}\land \d x^{\nu_1}\land \d x^{\mu_2}\land \d x^{\nu_2}\land \dots\land \d x^{\mu_k}\land \d x^{\nu_k}
\fl{fPolinomialFormEQ}$$

These local expressions glue together (because of the transformation laws $\ShowLabel{CurvatureTransformationLawsEQ}$ and the $\Ad$-invariance of the polynomial $f$) to uniquely determine a global form on $\La^{2k}(P)$.
The same expression defines a $2k$-form on $M$ since for a specific connection its curvature coefficients depend on spacetime coordinates  $x$ alone.
We remark that it may happen (for large enough $k$) that such a form is identically zero.

From now on we shall denote by $f(\al_1, \dots,\al_p, \be^{k-p})=f(\al_1, \dots,\al_p, \be, \be\dots,\be)$, $0\le p\le k$, with the last argument $\be$ repeated $(k-p)$ times.
Accordingly, we have $f(F^k)=f(F, F, \dots, F)$, with $F$ repeated $k$ times.

\bs\PROPERTYn
{\sl the form $f(F^k)$ is closed.}
\ENDPROPERTY
\ms\PROOF
it follows from Bianchi identities $\na F=0$:
$$
\d (f(F^k))= \na (f(F^k))= k f(\na F, F^{k-1})=0
\fn$$
\ENDPROOF

\bs\PROPERTYn
{\sl the cohomology class [$f(F^k)$]$\in H^{2k}(M, \R)$ is independent of the connection $A$.}
\ENDPROPERTY
\ms\PROOF
let us consider two connections $A$ and $\bar A$ on $\calP$.
The difference $\al=A-\bar A$ is a well-defined global section of $\La^1(P; \gotg)$
due to the affine structure on the bundle of connections $\calC(P)$, which is modelled
precisely on $\La^1(P; \gotg)$.
Let $\om_s= \bar A + s\al$, $s\in\R$ be the so-called {\it interpolating connection}
(again notice that the affine structure on $\calC(P)$ ensures that $\om_s$ is in fact a
$1$-parameter family of global connections on $\calP$).
Let us denote by $\Om_s$ the corresponding curvature, given by:
$$
 \Om_s=\bar F+s\bar\na\al+s^2\>(\al^2)
\fn$$
where, following the notation of Appendix $\AppendixTapia$, we set
$$
 \eqalign{
  &\Om_s := \d\om_s + \hbox{$1\over 2$}[\om_s,\om_s]\cr
  &\bar F := \d\bar A + \hbox{$1\over 2$}[\bar A,\bar A]\cr
}
\qquad
 \eqalign{
  &\bar \na\al :=\d \al + [\bar A, \al]\cr
  &(\al)^2 :=\hbox{$1\over 2$}\>[\al,\al]\cr
}
\fn$$

Moreover one easily obtains:
$$
{\d\over \d s}\Om_s= \bar\na \al  + s [\al,\al]=: \na^s \al
\fn$$
where $\na^s \al$ is the exterior covariant differential induced on $\La^1(P; \gotg)$ by the connection $\om_s$.

Thus we have:
$$
\eqalign{
f(F^k)- f(\bar F^k)=&\int_0^1 {\d\over \d s} f(\Om_s^k)\d s=
k\int_0^1  f\left({\d\over \d s}\Om_s^k,\Om_s^{k-1}\right)\d s=\cr
=&
k\int_0^1  f(\na^s\al,\Om_s^{k-1})\d s=
k\int_0^1  \na^s f(\al,\Om_s^{k-1})\d s=\cr
=&
k\int_0^1  \d f(\al,\Om_s^{k-1})\d s=
\d\left(k\int_0^1  f(\al,\Om_s^{k-1})\d s\right)\cr
}
\fn$$
It then follows $[f(F^k)]=[f(\bar F^k)]$.
\ENDPROOF
\bs
Let us now set $\tf(A, \bar A):=k\int_0^1  f(\al,\Om_s^{k-1})\d s$.
For two specific connections $A$ and $\bar A$ this is a $(2k-1)$-form on $P$, though the same expression defines a $(2k-1)$-form on $M$ as well (which will be denoted below by $\Tf(A, \bar A)$).
Given a background reference connection $\bar A$ we are then able to single out
a canonical representative $f(\bar F^k)+\d \tf(A, \bar A)$ for the cohomology class $[f(F^k)]$.

Locally on $P$, once a trivialization is chosen, we can set $\bar A=0$ (which of course
has no global meaning) and obtain
$$
f(F^k)= \d\> \tf(A, 0)\equiv  \d\> \tf(A)
\fn$$
which is a local potential (on $M$) for the form $f(F^k)$.

Notice that if $m=2k-1$ is odd both $f(F^k)$ and $f(\bar F^k)$ are zero and $\tf(A, \bar A)$ is a global $m$-form over $P$;
it is also transgressive, i.e.\ $\tf(A, \bar A)= \pi^\ast \Tf(A, \bar A)$ for some
$m$-form over $M$, $\Tf(A, \bar A)$ having the same local expression of $\tf(A, \bar A)$.
We shall see in the next Section that both Lagrangians $\Tf(A, \bar A)$ and
$\Tf(A)=\Tf(A, 0)$ produce the same global field equations (even if the second is just a family of local Lagrangians).
As a consequence  $\Tf(A, \bar A)$ is by construction the globalization of $\Tf(A)$ obtained by introducing a reference vacuum state.

\bs\PROPERTYn
{\sl the form $f(F^k)$ is transgressive.}
\ENDPROPERTY
\ms\PROOF
it follows immediately from the local expression $\ShowLabel{fPolinomialFormEQ}$.
The form $f(F^k)$ is fiberwise constant, hence projectable onto a form $\f(F^k)$ on $M$
given by the same local expression as $\ShowLabel{fPolinomialFormEQ}$ (we stress that the connection
has been fixed so that $F^A_{\mu\nu}(x)$ are fuctions of the variables $x$ only).
One trivially has $f(F^k)= \pi^\ast \f(F^k)$.
\ENDPROOF

\NewSection{Variational Properties of Chern-Simons Lagrangians}

From now on let us set $m=2k-1$.
The form  $\tf(A, \bar A)$ on $P$ can be canonically regarded as the evaluation of a horizontal form on
$J^1\calC(P)$ along the $1$-jet prolongation of a section $A$ of the bundle $\calC(P)$ of connections.
By an abuse of notation we shall consider the global Lagrangian (see also \ref{Izaurieta} and references quoted therein)
$$
L_{A\bar A}^{{}_{(k)}}= \Tf(A, \bar A)= k\int_0^1  f(\al,\Om_s^{k-1})\>\d s
\fl{CovariantLagrangianEQ}$$
and the local Lagrangians
$$
L_A^{{}_{(k)}}=\Tf(A)=k\int_0^1  f(A, (sF +\hbox{$1\over 2$} s(s-1)[A, A])^{k-1})\>\d s
\fn$$
where $[A, A]= c_{BC}{}^A A^B_\mu A^C_\nu\>T_A\otimes \d x^\mu \land \d x^\nu$ (which of course makes sense only in a trivialization).
If the trivialization is changed the connection transforms according to its
affine rules $\ShowLabel{ConnectionTransformationEQ}$ so that the polynomial $\Tf(A)$ (and consequently the Lagrangian $L_A^{{}_{(k)}}$) is not global.

The first variation formula for the Lagrangian $L_{A\bar A}^{{}_{(k)}}$ can be obtained:
$$
\de L_{A\bar A}^{{}_{(k)}}=\E(L_{A\bar A}^{{}_{(k)}}) + \d \F(L_{A\bar A}^{{}_{(k)}})
\fn$$
where we set
$$
\cases{
\E(L_{A\bar A}^{{}_{(k)}})= k f(\de A, F^{k-1}) - kf(\de \bar A, \bar F^{k-1})\cr
\F(L_{A\bar A}^{{}_{(k)}})=k(k-1) \int_0^1  f(\de \om_s, \al, \Om^{k-2}_s) \> \d s\cr
}
\fl{CovariantFieldEquations}$$
A detailed derivation of these results can be found below in Appendix $\AppendixFE$.

From these results one can obtain as a local special case the first variation formula for the local Lagrangians $L_A^{{}_{(k)}}$
$$
\de L_{A}^{{}_{(k)}}=\E(L_{A}^{{}_{(k)}}) + \d \F(L_{A}^{{}_{(k)}})
\fn$$
by simply inserting $\bar A=0$ into the equations \ShowLabel{CovariantFieldEquations}, i.e.\
$$
\cases{
\E(L_{A}^{{}_{(k)}})= k f(\de A, F^{k-1}) \cr
\F(L_{A}^{{}_{(k)}})=k(k-1) \int_0^1  f(s\de A, A, (s F + s(s-1) (A^2))^{k-2}_s) \>\d s =\cr
\qquad\quad= \sum_{i=0}^{k-2}{(-1)^i k!(k-1)!\over (k-i-2)!(k+i)!} \> f(\de A, A, (A^2)^i, F^{k-i-2})
}
\fn$$
We stress (referring again to the Appendix $\AppendixFE$) that specializing $k=2, 3, \dots$ one obtains the standard results obtained in any odd dimension $m=3, 5, \dots$ (see \ref{Silvio}).

We can now apply the results of \ref{Tapia} (in their form summarized in Appendix $\AppendixTapia$)
in order to split the covariant Lagrangian $L_{A\bar A}^{{}_{(k)}}$ as
$$
L_{A\bar A}^{{}_{(k)}}= L_A^{{}_{(k)}}- L_{\bar A}^{{}_{(k)}} + \d \De^{{}^{(k)}}
\fn$$
where $\De^{{}^{(k)}}$ is the form
$$
\De^{{}^{(k)}}= -\sum_{i=0}^{k-2} {(-1)^i k! (k-1)!\over (k-i-2)!(k+i)!} \int ^1_0 f(\al, \om_s, (\om_s^2)^i,(\Om_s)^{k-i-2} ) \> \d s
\fn$$

\NewSection{Conserved Quantities}

Both $\al$ and $\Om_s$ are sections of $\La^k(P;\gotg)$ (for $k=1$ and $k=2$, respectively) and
hence they transform according to $\ShowLabel{CurvatureTransformationLawsEQ}$ under principal automorphisms of the principal bundle $P$ (\ref{BFF}, \ref{BFFa}, \ref{AFR}, \ref{Stasheff}).
Being $f$ $\Ad$-invariant the Lagrangian $L_{A\bar A}^{{}_{(k)}}$ is $\Aut(P)$-covariant.
As a consequence (as one can also check directly by expanding both handsides and Lie derivatives) the following covariance identity holds true
$$
k\int^1_0 f(\Lie_\Xi \al, \Om_s^{k-1})\>\d s + k(k-1)\int^1_0 f(\al, \Lie_\Xi\Om_s, \Om_s^{k-2})\>\d s=
\d(\xi \ip L_{A\bar A}^{{}_{(k)}})
\fl{IdentitaCovarianzaEQ}$$
where we set $\Xi$ for the generator of infinitesimal gauge transformation which projects over the spacetime vector field $\xi$ and $\ip$ denotes the contraction of a vector field with an $m$-form.

Notice that we can use the connection $\om_s$ ($\bar A$, respectively) to define the vertical part of $\Xi$
$$
(\xi_{(V)}^s)^A = \Xi^A +(\om_s)^A_\mu \xi^\mu
\qquad\quad
[(\bar\xi_{(V)})^A = \Xi^A +\bar A^A_\mu \xi^\mu\hbox{, respectively}]
\fn$$
The vertical part of a vector field transforms with the adjoint ($\Ad$) representation, so that we can
define a section in $\La^0(P;\gotg)$ by setting
$$
\xi_{(V)}^s = (\xi_{(V)}^s)^A \>T_A\otimes\one
\fn$$
Let us also introduce the following notation for the contraction of the curvature along a spacetime vector field,
which turns out to be a section of $\La^1(P;\gotg)$
$$
\xi\ip\Om_s= \xi^\la(\Om_s)^A_{\la\mu}\>T_A\otimes \d x^\mu
\fn$$
and the $0$-form
$$
\xi\ip \al= \xi^\la \al^A_{\la}\>T_A\otimes \one
\fn$$
Using this notation we can express the Lie derivative of the connection as
$$
\Lie_\Xi \om_s= \xi\ip\Om_s + \na^s \xi^s_{(V)}
\fn$$

The N\"other current is easily obtained from $\ShowLabel{IdentitaCovarianzaEQ}$ as
$$
\eqalign{
\calE(L_{A\bar A}^{{}_{(k)}}, \Xi)=& k(k-1)\int^1_0 f(\Lie_\Xi\om_s, \al , \Om_s^{k-2})\>\d s
-k \int^1_0 \xi \ip f(\al, \Om_s^{k-1})\>\d s=\cr
=&
k(k-1)\int^1_0 f(\xi\ip\Om_s + \na^s \xi^s_{(V)}, \al , \Om_s^{k-2})\>\d s
-k \int^1_0 f(\xi \ip \al, \Om_s^{k-1})\>\d s+\cr
&\qquad+k(k-1) \int^1_0 f(\al, \xi \ip \Om_s, \Om_s^{k-2})\>\d s=\cr
=&
k(k-1)\int^1_0 f(\na^s \xi^s_{(V)}, \al , \Om_s^{k-2})\>\d s
-k \int^1_0 f(\xi \ip \al, \Om_s^{k-1})\>\d s\cr
}
\fn$$
This N\"other current can be covariantly integrated by parts
$$
\calE(L_{A\bar A}^{{}_{(k)}}, \Xi)= \tilde \calE(L_{A\bar A}^{{}_{(k)}}, \Xi) + \d \calU(L_{A\bar A}^{{}_{(k)}}, \Xi)
\fn$$
to define the {\it superpotential}
$$
\calU(L_{A\bar A}^{{}_{(k)}}, \Xi)= k(k-1) \int^1_0 f(\xi^s_{(V)}, \al, \Om_s^{k-2})\>\d s
\fl{CSSuperpotentialEQ}$$
and the {\it reduced current}
$$
\tilde \calE(L_{A\bar A}^{{}_{(k)}}, \Xi)=
-k(k-1) \int^1_0 f(\xi^s_{(V)}, \na^s\al, \Om_s^{k-2})\>\d s -k \int^1_0 f(\xi \ip \al, \Om_s^{k-1})\>\d s
\fn$$
One can directly check that the reduced current vanishes on-shell by simply expanding
$\Om_s= sF+(1-s)\bar F -s(1-s)(\al^2)$, $\xi^s_{(V)}=\bar \xi_{(V)} + s\xi\ip\al$,
$\na^s\al=\bar \na\al + s[\al, \al]\equiv {\d \Om_s\over\d s}$ and then by integrating by parts with respect to the derivative ${\d \Om_s\over\d s}$.
The result factorizes through field equations.

The superpotential $\ShowLabel{CSSuperpotentialEQ}$ can be specialized to the case $k=2$ to obtain
$$
\calU(L_{A\bar A}^{{}_{(2)}}, \Xi)= f(\xi_{(V)}+ \bar \xi_{(V)}, \al)
\fn$$
as already computed in $\ref{AFR}$.

We remark that the non-covariant Lagrangians are not invariant with respect to gauge transformations.
Only the total Lagrangian $L_{A\bar A}^{{}_{(k)}}=L_A^{{}_{(k)}}-L_{\bar A}^{{}_{(k)}} + \d\De^{{}^{(k)}}$ is covariant. N\"other theorem
and superpotential theory apply to the covariant Lagrangian only. In other words, despite the superpotential
$\calU(L_{A\bar A}^{{}_{(k)}}, \Xi)$ receives contributions from $L_A^{{}_{(k)}}$, $L_{\bar A}^{{}_{(k)}}$ and $\De^{{}^{(k)}}$, the corresponding splitting
of the superpotential $\calU(L_{A\bar A}^{{}_{(k)}}, \Xi)= \calU(L_{A}^{{}_{(k)}}, \Xi)- \calU(L_{\bar A}^{{}_{(k)}}, \Xi)+ \calU(\d\De^{{}^{(k)}}, \Xi)$
is not endowed with a fundamental meaning, each term depending on the trivialization (or equivalently on the gauge fixing).

For example, in the case $k=2$ the contributions are
$$
\cases{
\calU(L_{A}^{{}_{(2)}}, \Xi)= f(A, \xi_{(V)})\cr
\calU(L_{\bar A}^{{}_{(2)}}, \Xi)= f(\bar A, \bar\xi_{(V)})\cr
 \calU(\d\De^{{}^{(2)}}, \Xi)=f(A, \bar\xi_{(V)})- f(\bar A, \xi_{(V)})\cr
}
\fn$$
none of which is gauge invariant whilst their sum is.

\NewSection{BF Theory}

As suggested in \ref{BF} we can change variables in the space of
fields so to write Chern-Simons theory in a form which can be later
specialized
as a BF theory (see \ref{BFa}, \ref{BFb}, \ref{BFc}, \ref{BFd}, \ref{Montesinos}), i.e.~a
gauge natural theory for a connection and a section of
$\La^1(P;\gotg)$, see \ref{BFFa}.

The $3$-dimensional case has been already dealt with in \ref{BF}.
Here we are able to generalize our previous results to the case of
any odd dimension. Let us first define an {\it average connection}
$$
\tilde \om:= \om_\si=\si A + (1-\si) \bar A \fn$$ with $\si\in[0,1]$
and the {\it relative field} $\al=A-\bar A$. We shall use these two
fields $(\tilde \om, \al)$ as fundamental fields in place of the
original fields $A$ and $\bar A$. This provides a one-parameter
family of mutally equivalent topological field theories, the so-called BF
type theories.

The converse transformation is
$$
\cases{ A= \tilde \om +(1-\si) \al\cr \bar A=\tilde \om -\si\al\cr }
\fn$$
The interpolating connection $\om_s$ can be also expressed in
terms of $\tilde \om$ and $\al$ as
$$
\om_s= \tilde \om +(s-\si)\al
\fn$$
so that we obtain
$$
\Om_s= \tilde\Om + (s-\si)\tilde \na\al + (s-\si)^2(\al^2)
\fl{BFInterpolatingCurvatureEQ}$$ where $\tilde \Om$ is the
curvature of $\tilde \om$. Hence we can express the Lagrangian as
$$
L_{A\bar A}^{{}_{(k)}}= k\int^1_0 f(\al,\Om_s^{k-1})\>\d s
\fn$$
where $\Om_s$ is expressed in terms of the fundamental fields and their derivatives as in $\ShowLabel{BFInterpolatingCurvatureEQ}$.

For $k=2$ we obtain the results of \ref{BF}, i.e.
$$
L_{A\bar A}^{{}_{(2)}}=f(\al, 2\tilde\Om +(1-2\si)\tilde\na\al +\hbox{$1\over 3$}
(1+3\si^2-3\si)[\al,\al])
\fn$$
We stress that here $\si$ has to be interpreted as a parameter to be specialized
{\it a priori}.

Field equations can be easily obtained by
$$
\eqalign{
\E(L_{A\bar A}^{{}_{(k)}})=& kf(\de A, F^{k-1})-kf(\de\bar A, \bar F^{k-1})=\cr
=&kf(\de \tilde\om, F^{k-1}-\bar F^{k-1}) + kf(\de \al, (1-\si )F^{k-1}+\si\bar F^{k-1})
}
\fn$$
where $F$ and $\bar F$ are meant to be expressed by
$$
\cases{
F= \tilde \Om +(1-\si) \tilde\na\al +(1-\si)^2 (\al^2)\cr
\bar F=\tilde \Om -\si \tilde\na\al +\si^2 (\al^2)\cr
}
\fn$$
Again for $k=2$ we obtain the known results
$$
\E(L_{A\bar A}^{{}_{(2)}})=2 f(\de \tilde\om, \tilde\na\al -(2\si-1)(\al^2))
+ 2f(\de \al, \tilde\Om +\si(1-\si)(\al^2)+[\tilde\na\al -(2\si-1)(\al^2)])
\fn$$

The superpotential can be recast as
$$
\calU(L_{A\bar A}^{{}_{(k)}}, \Xi)=k(k-1) \int^1_0 f(\xi^s_{(V)}, \al, \Om_s^{k-2})\>\d s
\fn$$
which for $k=2$ specializes to
$$
\calU(L_{A\bar A}^{{}_{(2)}}, \Xi)= f(\tilde\xi_{(V)}+(1-2\si)\xi\ip\al, \al)
\fn$$
Here $\tilde\xi_{(V)}$ denotes the vertical part of the symmetry generator $\Xi$
with respect to the connection
$\tilde \om$.

\NewSection{Conclusions and Perspectives}

We have here extended the formulation of covariant Chern-Simons theories to an arbitrary odd dimension and
investigated their relation with BF theories.
The introduction of a vacuum reference field $\bar A$ allows  to define a global gauge covariant Lagrangian depending on two dynamical fields $A$ and $\bar A$ both obeing the (global) field equations defining Chern-Simons models.
We stress that the reference field is not to be interpreted as a background (in the sense in which Minkowski spacetime is a background in particle field theory) and it can be given a suitable physical interpretation (see \ref{Augmented}).

The resulting global Lagrangian for Chern-Simons theory is quite similar to the covariant first order Lagrangian
for GR
(see \ref{Augmented}, \ref{Root}, \ref{Cavalese}) despite the fact that it is obtained by relying on quite different motivations (see \ref{Silvio}).

The understanding of the geometric relation between conservation laws for gauge-like Chern-Simons theories and
Chern-Simons gravity is, at least to some extent, still partially obscure.
In fact, these two models are gauge natural field theories for two different structure groups.
 We begin to understand this relation in some example (see, e.g., \ref{Lovelock}, \ref{AFR}), but
there is yet no general framework able to deal with these sort of transformations in full generality.
This will form the subject of future investigations.

\NewAppendix{\AppendixTapia}{Lagrangian Splittings Induced by Homotopies in the Space of Connections}

We shall review the results presented in $\ref{Tapia}$ here suitably adapted to our case,i.e.~a field theory on the bundle of connections.

Let us first introduce some notation. Let $A$ and $\bar A$ be two connections on a principal bundle $P$ and
let us set $\al=A-\bar A$, which is a section of the model vector bundle $\La^1(P; \gotg)$. We define
$$
\om_s:= \bar A + s\al= sA+(1-s)\bar A
\fn$$
for the {\it interpolating connection}, $s$ being any real number.

Let $\Om_s:= \d \om_s + \hbox{$1\over 2$}[\om_s,\om_s]$ denote the curvature of the interpolating connection $\om_s $. We stress that $\Om_s$ differs from the convex interpolation of the curvatures
$(F, \bar F)$ of the connections $(A, \bar A)$, respectively. In fact, one has
$$
\eqalign{
\Om_s= & \bar F + s \bar \na \al + s^2(\al^2)=\cr
=& s F +(1-s) \bar F + s(s-1) (\al^2)\cr
}
\fn$$
where $\bar \na \al := \d \al + [\bar A, \al]$ denotes the {\it exterior gauge-covariant differential} of $\al$
with respect to the connection $\bar A$.

Let us consider a first order (not necessarily gauge covariant or global) Lagrangian $L=L(A, F)$.
Its first variation formula reads as
$$
\eqalign{
\de L=& {\del L\over \del A} \de A + {\del L\over \del F} \de F =
{\del L\over \del A} \de A + {\del L\over \del F} \na\de A=\cr
=& <\E(L)\>|\> \de A> + \d <\F(L)\>|\>\de A>\cr
}
\fl{FVFTapiaEQ}$$
where $\na \de A$ denotes the exterior gauge-covariant differential with respect to the connection $A$
and we set
$$
\left\{
\eqalign{
&<\E(L)\>|\> \de A>= \left({\del L\over \del A} - \na  {\del L\over \del F}\right)\de A\cr
&<\F(L)\>|\> \de A>=  {\del L\over \del F^A_{\mu\nu}} \de A^A_\mu \>\d \si_\nu\cr
}\right.
\fn$$
where $\d \si_\nu$ is the canonical local basis for $(m-1)$-forms on $M$ induced by coordinates.

Let us now consider the auxiliary Lagrangian
$$
\tilde \La:=\int_0^1 {\d L_s\over \d s}\>\d s
\fl{AuxiliaryLagrangianEQ}$$
where we set $L_s = L(\om_s, \Om_s)$.
This auxiliary Lagrangian can be computed explicitely as
$$
\tilde \La=L_1- L_0
\fn$$
by direct integration of $\ShowLabel{AuxiliaryLagrangianEQ}$ and by using the first variation formula $\ShowLabel{FVFTapiaEQ}$
$$
\eqalign{
\tilde \La= \int^1_0 \left( {\del L\over \del \om_s} {\d \om_s\over \d s}
+ {\del L\over \del \Om_s} {\d \Om_s\over \d s}\right)\d s=
\int^1_0 \left( {\del L\over \del \om_s}\al
+ {\del L\over \del \Om_s} \na^s \al \right)\d s
}
\fn$$
where we set $\na^s \al:=\bar \na \al +s [\al,\al]$ for the exterior gauge-covariant differential with respect to the interpolating connection $\om_s$.
By a gauge-covariant integration by part and comparing with first the variation formula $\ShowLabel{FVFTapiaEQ}$ we obtain
$$
\tilde \La=  \int^1_0 <\E(L_s)\>|\> \al> \> \d s  + \d \int^1_0 <\F(L_s)\>|\> \al> \> \d s
\fn$$

By finally defining
$$
\cases{
\La= \int^1_0 <\E(L_s)\>|\> \al> \> \d s\cr
\De= -\int^1_0 <\F(L_s)\>|\> \al> \> \d s
}
\fn$$
one easily obtains the Lagrangian splitting
$$
\La= L_1- L_0 + \d \De
\fl{LagrangianSplittingEQ}$$

Let us remark the following: $(i)$ $L_1= L(A, F)$ and $L_0= L(\bar A, \bar F)$; $(ii)$  $\La$ has the same field equations
of $L$ (one for $A$ and one for $\bar A$); $(iii)$ $\La$ is obtained by integrating the field equations of the
original Lagrangian $L$ along the interpolating homotopy.

This method of producing an equivalent Lagrangian which splits is particularly effective for Chern-Simons theory.
In fact, the method does not rely on any assumption of gauge covariance on the original Lagrangian (so that it can be applied also to the Chern-Simons non-covariant Lagrangians).
The resulting Lagrangian $\La$ relies on field equations (which for Chern-Simons theory {\it are} in fact gauge covariant) and it is global and gauge covariant by construction.

As we shall prove in the Appendix $\AppendixFE$, field equations for the non-covariant Chern-Simons Lagrangians are in the form
$$
<\E(L_A^{{}_{(k)}})\>|\> \de A>= kf(\de A, F^{k-1})
\fn$$
which produces the Lagrangian
$$
\La = k \int^1_0 f(\al, \Om_s^{k-1}) \> \d s
\fn$$
which, in turn, coincides with the covariant Chern-Simons Lagrangian $L_{A\bar A}^{{}_{(k)}}$ given by
$\ShowLabel{CovariantLagrangianEQ}$.

The ensuing splitting \ShowLabel{LagrangianSplittingEQ} reads now as
$$
L_{A\bar A}^{{}_{(k)}}= L_A^{{}_{(k)}}- L_{\bar A}^{{}_{(k)}} + \d \De^{{}^{(k)}}
\fl{CSLagrangianSplittingEQ}$$
with the form
$$
\De^{{}^{(k)}}= -\sum_{i=0}^{k-2} {(-1)^i k! (k-1)!\over (k-i-2)!(k+i)!} \int ^1_0 f(\al, \om_s, (\om_s^2)^i,(\Om_s)^{k-i-2} ) \> \d s
\fl{GeneralCovariantLagrangianSplittingEQ}$$
where we used the explicit formula for $\F(L_A^{{}_{(k)}})$ computed in Appendix $\AppendixFE$ below.

The main reason which justifies the use of the results of \ref{Tapia} is that they allow to explicitely compute the divergence term $\De$
in the splitting  $\ShowLabel{CSLagrangianSplittingEQ}$, which is in fact rather tedious to be computed directly in the generic dimension case; see \ref{Silvio}.

\NewAppendix{\AppendixFE}{Field Equations for the Covariant Lagrangian}

We shall here compute field equations for the covariant Chern-Simons Lagrangian $L_{A\bar A}^{{}_{(k)}}$
introduced in $\ShowLabel{CovariantLagrangianEQ}$.
Of course one could simply notice that because of the Lagrangian decomposition $\ShowLabel{LagrangianSplittingEQ}$ field equations  of $L_{A\bar A}^{{}_{(k)}}$ have to coincide with field equations of the non-covariant Lagrangians $L_{A}^{{}_{(k)}}$, which would simplify the computation considerably.
However, we shall here compute field equations for the covariant Lagrangian directly, in order to become more familiar with the tricks that are used in this paper  for conservation laws and also in order to definitely clarify the relation between
the trasgression technique we used when defining the covariant Lagrangian $L_{A\bar A}^{{}_{(k)}}$ and the
homotopic technique used in \ref{Tapia} in order to construct the Lagrangian splitting $\ShowLabel{LagrangianSplittingEQ}$; one needs in fact to know that the covariant Lagrangian $L_{A\bar A}^{{}_{(k)}}$ is precisely in the form $\ShowLabel{CovariantLagrangianEQ}$.
Moreover, we compute also the Poincar\'e-Cartan morphism which will be used to express N\"other currents and conservation laws.

We start by considering the Lagrangian variation
$$
\eqalign{
\de L_{A\bar A}^{{}_{(k)}}=& k\int^1_0 f(\de \al, \Om_s^{k-1})\>\d s
+  k(k-1)\int^1_0 f( \al, \de\Om_s, \Om_s^{k-2})\>\d s=\cr
=& k\int^1_0\left[ f(\de \al, \Om_s^{k-1}) + (k-1) f( \al, \na^s\de\om_s, \Om_s^{k-2})\right]\>\d s=\cr
=&k\int^1_0\left[  f(\de \al, \Om_s^{k-1}) +  (k-1) f(\na^s \al, \de \om_s, \Om_s^{k-2})\right]\>\d s+\cr
&-k(k-1)\int^1_0 \d f(\al, \de \om_s, \Om_s^{k-2})\>\d s=\E(L_{A\bar A}^{{}_{(k)}}) +\d \F(L_{A\bar A}^{{}_{(k)}}) \cr
}
\fn$$
where Bianchi identities $\na^s\Om_s=0$ were used and we have set
$$
\cases{
\E(L_{A\bar A}^{{}_{(k)}})= k\int^1_0\left[  f(\de \al, \Om_s^{k-1})
+  (k-1) f(\na^s \al, \de \om_s, \Om_s^{k-2})\right]\>\d s\cr
\F(L_{A\bar A}^{{}_{(k)}})=k(k-1)\int^1_0  f(\de \om_s, \al, \Om_s^{k-2})\> \d s\cr
}
\fn$$

Let us now expand the Euler-Lagrange morphism $\E(L_{A\bar A}^{{}_{(k)}})$ as follows:
$$
\eqalign{
\E(L_{A\bar A}^{{}_{(k)}})=& k\int^1_0\big[  f(\de \al, \Om_s^{k-1})
+  (k-1) f(\de \bar A, \na^s \al,  \Om_s^{k-2})+\cr
&+  (k-1) f(\de \al, \Om_s -\bar F +s^2(\al^2), \Om_s^{k-2})\big]\>\d s=\cr
=& k^2\int^1_0 f(\de A,\Om_s-a\bar F+as^2(\al^2),\Om_s^{k-2}) \>\d s+\cr
&-k^2\int^1_0 f(\de \bar A,\Om_s-a \na^s\al -a\bar F + as^2(\al^2),\Om_s^{k-2}) \>\d s
\cr}
\fn$$
where $a=\hbox{$k-1\over k$}$.

\bs\THEOREMl{ELTH}
The following holds:
$$
 k^2\int^1_0 f(\de A,\Om_s-a\bar F+as^2(\al^2),\Om_s^{k-2}) \>\d s= kf(\de A, F^{k-1})
\fl{ThesisEQ}$$
\ENDTHEOREM
\ms\PROOF
first expand $\Om_s= s F + (1-s)\bar F +s(s-1)(\al^2)$, then use the power formula and recall that
$$
\int_0^1 s^m(1-s)^n\>\d s= {m!n!\over (m+n+1)!}
\fn$$
as one can prove by a multiple integration by parts.
The left hand side of equation $\ShowLabel{ThesisEQ}$ can thus be recast as:
$$
\eqalign{
k^2\Big[&
\sum_{i=0}^{k-2}\sum_{j=0}^{k-i-2} (-1)^i \binomial{k-2}{i}\binomial{k-i-2}{j}{(k-j-1)!(i+j)!\over (k+i)!}+\cr
&+{1\over k}\sum_{i=0}^{k-2}\sum_{j=1}^{k-i-1} (-1)^i \binomial{k-2}{i}\binomial{k-i-2}{j-1}{(k-j-1)!(i+j-1)!\over (k+i-1)!}+\cr
&-\sum_{i=0}^{k-2}\sum_{j=1}^{k-i-1} (-1)^i \binomial{k-2}{i}\binomial{k-i-2}{j-1}{(k-j)!(i+j-1)!\over (k+i)!}+\cr
&-{2k-1\over k}\sum_{i=1}^{k-1}\sum_{j=0}^{k-i-1} (-1)^i \binomial{k-2}{i-1}\binomial{k-i-1}{j}{(k-j)!(i+j-1)!\over (k+i)!}+\cr
&+\sum_{i=1}^{k-1}\sum_{j=0}^{k-i-1} (-1)^i \binomial{k-2}{i-1}\binomial{k-i-1}{j}{(k-j-1)!(i+j-1)!\over (k+i-1)!}
\Big]\cdot\cr
&\qquad\cdot f(\de\al, F^{k-i-j-1}, \bar F^j, (A^2)^i)
}
\fn$$
The first term is the only one which contributes to the term with $i=0$ and $j=0$ and it produces exactly $kf(\de A, F^{k-1})$ while all the other contributions cancel away.
\ENDPROOF

\bs\COROLLARYn one has
$$
 k^2\int^1_0 f(\de \bar A,\Om_s-a \na^s\al -a\bar F + as^2(\al^2),\Om_s^{k-2})= kf(\de \bar A, \bar F^{k-1})
\fn$$
\ENDCOROLLARY
\PROOF
just change the integration variable $\si=1-s$ and use theorem $\ShowLabel{ELTH}$.
\ENDPROOF

\bs
Hence we have obtained
$$
\E(L_{A\bar A}^{{}_{(k)}})=kf(\de A, F^{k-1})-kf(\de \bar A, \bar F^{k-1})
\fl{GeneralCovariantFieldEquationEQ}$$

As far as the Poincar\'e-Cartan morphism $\F(L_{A\bar A}^{{}_{(k)}})$ is concerned we expand
$\Om_s= s F + (1-s)\bar F +s(s-1)(\al^2)$ as above and obtain
$$
\eqalign{
\F(L_{A\bar A}^{{}_{(k)}})=&k(k-1)\int^1_0  f(\de \om_s, \al, \Om_s^{k-2})\> \d s=\cr
=&k (k-1) \sum_{i=0}^{k-2} \sum_{j=0}^{k-i-2}(-1)^i  \binomial{k-2}{i}\binomial{k-i-2}{j}\cdot\cr
&\cdot
  \Bigg[{(k-j-1)!(i+j)!\over (k+i)!} f(\de A, \al, (\al^2)^i, \bar F^j, F^{k-i-j-2}+\cr
&\quad  +{(k-j-2)!(i+j+1)!\over (k+i)!} f(\de \bar A, \al, (\al^2)^i, \bar F^j, F^{k-i-j-2}
  )\Bigg]
}
\fl{CovariantPCMorphismEQ}$$
Then specializing $\bar A=0$ in $\ShowLabel{CovariantPCMorphismEQ}$ we obtain the Poincar\'e-Cartan morphism of the
local Lagrangians:
$$
\eqalign{
\F(L_{A}^{{}_{(k)}})=&
\sum_{i=0}^{k-2} (-1)^i  {k!(k-1)!\over (k-i-2)!(k+i)!} f(\de A, A, (A^2)^i, F^{k-i-2})\cr
}
\fn$$
which agrees with the prescription $\ShowLabel{GeneralCovariantLagrangianSplittingEQ}$.

Finally we can specialize everything to the specific cases $k=2,3$ (see \ref{Silvio}).

For $k=2$ we have the covariant Lagrangian
$$
L_{A\bar A}^{{}_{(2)}}= 2\int^1_0 f(\al,\Om_s)\>\d s = f(\al, F) + f(\al, \bar F) -\hbox{$1\over 3$} f(\al, \al^2)
\fn$$
the non-covariant Lagrangians
$$
L_A= f(A, F)-\hbox{$1\over 3$} f(A, A^2)
\fn$$
the field equations (see $\ShowLabel{GeneralCovariantFieldEquationEQ}$)
$$
\E(L_{A\bar A}^{{}_{(2)}})= 2 f(\de A, F) -2 f(\de \bar A, \bar F) =\E(L_{A}^{{}_{(2)}})- \E(L_{\bar A}^{{}_{(2)}})
\fn$$
the Poincar\'e-Cartan morphism (see $\ShowLabel{CovariantPCMorphismEQ}$)
$$
\F(L_{A\bar A}^{{}_{(2)}})= f(\de A, \al) + f(\de \bar A, \al)
\quad \then
\F(L_{A}^{{}_{(2)}})= f(\de A, A)
\fn$$
and finally the splitting $L_{A\bar A}^{{}_{(2)}}= L_{A}^{{}_{(2)}}- L_{\bar A}^{{}_{(2)}} + d\De^{{}^{(2)}}$ with (see $\ShowLabel{GeneralCovariantLagrangianSplittingEQ}$)
$$
\De^{{}^{(2)}} = - f(A, \bar A)
\fn$$

Analogously, for $k=3$ we have the covariant Lagrangian
$$
\eqalign{
L_{A\bar A}^{{}_{(3)}}=& 3\int^1_0 f(\al,\Om_s,\Om_s)\>\d s = f(\al, F^2) + f(\al, \bar F^2) + f(\al, F, \bar F)
+\cr
& +\hbox{$1\over 10$} f(\al, \al^2) -\hbox{$1\over 2$} f(\al, F, \al^2) -\hbox{$1\over 2$} f(\al, \bar F, \al^2)
}
\fn$$
the non-covariant Lagrangians
$$
L_A^{{}_{(3)}}= f(A, F^2)+\hbox{$1\over 10$} f(A, A^2, A^2) -\hbox{$1\over 2$} f(A, F, A^2)
\fn$$
the field equations (see $\ShowLabel{GeneralCovariantFieldEquationEQ}$)
$$
\E(L_{A\bar A}^{{}_{(3)}})= 3 f(\de A, F^2) -3 f(\de \bar A, \bar F^2) =\E(L_{A}^{{}_{(3)}})- \E(L_{\bar A}^{{}_{(3)}})
\fn$$
the Poincar\'e-Cartan morphism (see $\ShowLabel{CovariantPCMorphismEQ}$)
$$
\eqalign{
\F(L_{A\bar A}^{{}_{(3)}})=& f(\de A, \al, 2F +\bar F -\hbox{$1\over 2$} \al^2)
+ f(\de \bar A, \al, F +2\bar F -\hbox{$1\over 2$} \al^2)
\quad \then \cr
&\qquad\then\quad\F(L_{A}^{{}_{(3)}})= f(\de A, A, 2F -\hbox{$1\over 4$}[A, A])
}
\fn$$
and finally the splitting $L_{A\bar A}^{{}_{(3)}}= L_{A}^{{}_{(3)}}- L_{\bar A}^{{}_{(3)}} + d\De^{{}^(3)}$ with (see $\ShowLabel{GeneralCovariantLagrangianSplittingEQ}$)
$$
\De^{{}^(3)} = - f(A, \bar A, F+\bar F)
+\hbox{$1\over 4$} f(A, \bar A, [\al, \al])
+\hbox{$1\over 4$} f(A, \bar A, [A, \bar A])
\fn$$
One can check these results against \ref{Silvio}. Here they are computed by specializing general formulae
which are valid in any odd dimension, while they were there directly computed in those specific low dimensional cases, by means of {\it ad hoc} prescriptions for calculation.

\Acknowledgements

We wish to thank G. Allemandi, S. Mercadante and J. Stasheff for interesting discussions and comments.

This work is partially supported by GNFM-INdAM research project ``Metodi Geometrici in Meccanica
Classica, Teoria dei Campi e Termodinamica'' and by MIUR: PRIN 2003 on ``Conservation Laws and
Thermodynamics in Continuum Mechanics and Field Theories''.  We also acknowledge the contribution of
INFN (Iniziativa Specifica NA12) and the local research founds of Dipartimento di Matematica of Torino University.
One of us, A.B., is also partially supported by KBN-1P03B01828.
\ShowBiblio

\end